\documentclass[a4paper,11pt]{article}
\usepackage{jcappub} 

\pdfoutput=1 

\usepackage{jcappub} 

\usepackage[T1]{fontenc} 
\usepackage{graphicx}
\usepackage{epsfig}
\usepackage{rotate}
\usepackage{amsmath}
\usepackage{amssymb}
\usepackage{amsfonts}
\usepackage{multirow}
\usepackage{tablefootnote}
\usepackage{diagbox}
\usepackage{slashbox}
\usepackage{bm}
\usepackage{enumerate}
\usepackage{afterpage}
\usepackage{xcolor}
\usepackage{natbib, hyperref}
\usepackage{cancel}
\usepackage{amsmath,amssymb}

\def\be{\begin{equation}}
\def\ee{\end{equation}}
\def\bea{\begin{eqnarray}}
\def\eea{\end{eqnarray}}

\def\d{\mathrm{d}}

\allowdisplaybreaks

\title{\boldmath The Odd-Parity Part of the Observed Galaxy Trispectrum}

\author{Pritha Paul$^{1}$, Chris Clarkson$^{1,2}$ and Roy Maartens$^{2,3,4}$}
\affiliation{ $^{1}$School of Physics \& Astronomy, Queen Mary University of London, London E1 4NS, UK \\
$^2$Department of Physics \& Astronomy, University of the Western Cape, Cape Town 7535, South Africa\\
$^{3}$Institute of Cosmology \& Gravitation, University of Portsmouth, Portsmouth PO1 3FX, UK\\
$^4$National Institute for Theoretical and Computational Sciences (NITheCS), Cape Town 7535, South Africa
}

\emailAdd{p.paul@qmul.ac.uk}

\abstract{Recently the galaxy matter density $4$-point correlation function has been looked at to investigate parity violation in large scale structure surveys. The $4$-point correlation function is the lowest order statistic which is sensitive to parity violation, since a tetrahedron is the simplest shape that cannot be superimposed on its mirror image by a rotation. If the parity violation is intrinsic in nature, this could give us a window into inflationary physics. However, we need to exhaust all other contaminations before we consider them to be intrinsic. Even though the standard Newtonian redshift-space distortions are parity symmetric, the full relativistic picture is not. Therefore, we expect a parity-odd trispectrum when observing in redshift space. We calculate the trispectrum with the leading-order relativistic effects and investigate in detail the parameter space of the trispectrum and the effects of these relativistic corrections for different parameter values and configurations. We also look at different surveys and how the evolution and magnification biases can be affected by different parameter choices.}

\begin{document}
\maketitle
\flushbottom
\section{Introduction}
In the standard cosmological model, we expect large-scale structure statistics to be invariant under a parity transformation. If a parity violation is detected in the cosmological observables, it would point to physics which is beyond the standard model. Parity inversion in $3$-D is defined as reversing the sign of each spatial Cartesian coordinate axis at any event. Amongst the fundamental forces, only the weak interactions violate parity inversion. Conventionally, cosmological correlators are considered to be symmetric under a parity inversion. Despite that, there have been numerous studies searching of signatures of parity violation. In order to detect parity violation, we require observables that are sensitive to parity. Usually, vector and tensor quantities are the obvious choice. Research has mainly been focused on B-mode polarisation cross-correlation functions of the cosmic microwave background (CMB). The rotation of the CMB polarisation plane is often referred to as cosmic birefringence. Work has been done to constrain CMB birefringence \cite{Pogosian_2019, zagatti2024planckconstraintscosmicbirefringence}. Recently,  \cite{Minami_2020} searched for evidence of parity violation in the Planck 2018 data and found $2.4 \sigma$ evidence of a cosmic birefringence angle. This was updated by \cite{Eskilt_2022} which reported on a statistical significance of $3.6 \sigma$. Breaking of parity is also considered through terms that modify Einstein gravity, such as the Chern-Simons term, and whether they have left inflationary imprints \cite{Alexander_2009}. \cite{Bartolo_2017} considered parity violation in the polarisation of primordial gravitational waves, due to the Chern-Simons term coupled to an inflaton field. 

When looking for a parity-sensitive statistic in the scalar sector, such as the galaxy density field, it is harder to construct one. For example, the $2$-point correlation function is not sensitive to parity because a parity transformation is equivalent to a rotation. This also applies to the $3$-point correlation function. However recently, \cite{cahn2021test} proposed a novel method of testing parity invariance in the large-scale structure using the $4$-point correlation function ($4$PCF). The $4$PCF is represented by three separation vectors to three galaxies from the primary galaxy. This forms a tetrahedron. In $3$D, a tetrahedron cannot be superimposed on its mirror image, thus making it the lowest-order geometric shape that is sensitive to parity. As a consequence, in the scalar sector, the $4$PCF or the trispectrum is the lowest order statistic that can be used to detect parity violation. Following this, \cite{Hou_2023} also measured the parity-odd $4$PCF from BOSS data. They performed the analysis on DR$12$ LOWZ $(\bar{z} = 0.32)$  and DR$12$ CMASS $(\bar{z} = 0.57)$ and found $3.1 \sigma$ and $7.1 \sigma$ evidence respectively. Subsequently, \cite{Philcox_2022} performed a blind test on the parity-odd $4$PCF on the Baryonic Oscillation Spectroscopic (BOSS) CMASS sample of the Sloan Digital Sky Survey (SDSS) $\mathrm{III}$, finding a $2.9 \sigma$ evidence for a parity-odd $4$PCF in the range $20 - 160 h^{-1}\mathrm{Mpc}$. However, \cite{Philcox_2022} updated their results recalculating the covariance in the data, which brought down the evidence to $1.4 \sigma$ in the BOSS data set. Further, \cite{Krolewski:2024paz} and \cite{hewson2024spatialdistributionlargescalestructure}  did separate analyses and showed that there was no compelling evidence for parity violation in the BOSS data set. 

However, it is imperative that the $4$PCF or trispectrum is now investigated in more detail. So far, the power spectrum has been a key observable to probe the $\Lambda$CDM model. Over the past decade, there has been rapid advancement in large-scale structure surveys. Future and current galaxy surveys such as DESI, Euclid and SKAO are going to probe larger fractions of the sky. When comparing observational data to theory, it is becoming more important that we take observational projection effects into account. 
In 1987,  \cite{1987MNRAS.227....1K} showed that the observed galaxy number counts are not just sensitive to the underlying dark matter distribution. They are also affected by the peculiar velocity of the galaxies: since galaxies are measured in redshift space, the volume element constructed, which is a function of the observed redshift and observed angle, is different from the actual physical volume. Apart from the standard redshift space distortions (RSD), several other effects distort the number counts as well, since we are observing on our past light cone. These include lensing convergence, Doppler, Sachs-Wolfe, Integrated Sachs-Wolfe and time-delay effects. 

In addition to the Newtonian RSDs, we also have relativistic corrections that contain terms arising from Doppler terms and gravitational redshift. These relativistic distortions are suppressed by at least a factor of $\mathcal{H}/k$ in comparison to the standard RSD effects. The relativistic projection effects were computed at first order by [\citealp{Yoo_2009},\citealp{Yoo_2010},\citealp{Bonvin_2011}], subsequently extended to second  [\citealp{Bertacca:2014dra}, \citealp{DiDio:2014lka}, \citealp{Yoo:2014kpa}] and third order \cite{Dio_2019}. 

The derivation of the galaxy number counts shows us how changes in the density, velocity and metric result in a distortion of our past lightcone. For the power spectrum if we consider two galaxy populations, the terms that scale as $\mathcal{H}/k$ give rise to an imaginary part of the power spectrum \cite{McDonald_2009}, which induces an asymmetry \cite{Alam_2017, Cai_2017, Breton_2018}. This corresponds to the odd multipoles in the $2$PCF \cite{Bonvin_2014}. For the bispectrum and trispectrum, these effects come into play for a single galaxy survey \cite{Clarkson:2018dwn, Jeong_2020}. In this paper, we explore this parity-odd contribution to the trispectrum, which arises due to the above relativistic effects. These effects on the observed galaxy number counts shed light on the potential for galaxy surveys to further probe the geometry of our universe. Including these relativistic effects also shows how the perturbations in the density, peculiar velocity and metric distort the coordinate system and affect observables. The relativistic effects are important for tests of general relativity on these large scales and highlight the need for new observables that help to isolate these effects. This paper is follow up work from \cite{Paul_2024}. We give details behind the calculation of the trispectrum and investigate the parameters that can influence the parity-odd contributions at different scales. This paper is organised as follows. In section \ref{parityvio}, we give a brief overview of parity violation in large scale structure. In section \ref{reltrispec}, we introduce the relativistic trispectrum. In section \ref{thirdorderoverden} we give a general expression of the galaxy number overdensity and explicitly state the third order Newtonian and relativistic kernels that are required to calculate the trispectrum. Following that, in section \ref{numrel}, we construct the geometry for the trispectrum and present numerical results for the odd-parity contribution to the trispectrum. Appendix \ref{newtcoeff} and \ref{relcoeff} contains all the details of the coefficients for the Newtonian and the relativistic kernels.

\section{Parity violation in large scale structure} \label{parityvio}

For a density field $\delta(\boldsymbol{k})$, a parity transformation refers to $\boldsymbol{k} \rightarrow \boldsymbol{-k}$, resulting in $\delta(\boldsymbol{k}) \rightarrow \delta(\boldsymbol{-k})$. The reality condition holds for any real field: which implies $\delta(-\boldsymbol{k}) = \delta^*(\boldsymbol{k})$. Here $\delta^*$ is the complex conjugate of $\delta$. For any scalar field's N-point spectrum undergoing a parity transformation, we have: 
\begin{align} \label{parmap}
    F = \langle \delta(\boldsymbol{k}_1)\delta(\boldsymbol{k}_2)...\delta(\boldsymbol{k}_{\mathrm{N}-1})\delta(\boldsymbol{k}_{\mathrm{N}}) \rangle &\rightarrow \langle\delta(-\boldsymbol{k}_1)\delta(-\boldsymbol{k}_2)...\delta(-\boldsymbol{k}_{\mathrm{N}-1})\delta(-\boldsymbol{k}_{\mathrm{N}})\rangle \, , \\ \notag
    & = \langle \delta^*(\boldsymbol{k}_1)\delta^*(\boldsymbol{k}_2)...\delta^*(\boldsymbol{k}_{\mathrm{N}-1})\delta^*(\boldsymbol{k}_{\mathrm{N}}) \rangle = F^*.
\end{align}
From (\ref{parmap}), we can conclude that a parity transformation in Fourier space results in mapping the spectrum onto its complex conjugate. When isolating the odd-parity part of any N-point spectrum, we define it to be: 
\begin{align}
    F_{\mathrm{odd}} &= \frac{1}{2} \Big[\langle \delta(\boldsymbol{k}_1)\delta(\boldsymbol{k}_2)...\delta(\boldsymbol{k}_{\mathrm{N}-1})\delta(\boldsymbol{k}_{\mathrm{N}})\rangle -  \langle \delta^*(\boldsymbol{k}_1)\delta^*(\boldsymbol{k}_2)...\delta^*(\boldsymbol{k}_{\mathrm{N}-1})\delta^*(\boldsymbol{k}_{\mathrm{N}})\rangle\Big] \, ,\\
    &= i\mathrm{Im}\Big[F \Big].
\end{align}
Hence, we can see that the odd parity of any N-point correlation function is the imaginary part of the spectrum. For a real density field, the $3$-D power spectrum can be defined as  
\bea
P(\boldsymbol{k}_1) = \langle \delta(\boldsymbol{k}_1) \delta(\boldsymbol{k}_2) \rangle = \langle \delta(\boldsymbol{k}_1) \delta(-\boldsymbol{k}_1) \rangle = \langle \delta(\boldsymbol{k}_1) \delta^*(\boldsymbol{k}_1) \rangle = \langle |\delta(\boldsymbol{k}_1)|^2\rangle .
\eea
We can immediately see that the power spectrum is not sensitive to a parity transformation as it is purely real. This remains true also in redshift space. We can extend the analysis to the bispectrum. The bispectrum is given by: 
\bea
\langle \delta(\boldsymbol{k}_1) \delta(\boldsymbol{k}_2) \delta(\boldsymbol{k}_3) \rangle = (2 \pi)^3 \delta^D(\boldsymbol{k}_1 + \boldsymbol{k}_2 + \boldsymbol{k}_3) B(\boldsymbol{k}_1,\boldsymbol{k}_2,\boldsymbol{k}_3).
\eea
It is less clear for the bispectrum. However, we can see that the three wave vectors of the bispectrum should form a closed triangle due to the Dirac delta function. When a triangle is parity transformed in $3$-D, this is equivalent to a rotation. This tells us that the bispectrum is also not sensitive to parity transformation. However, when we observe in redshift space, we need to take the line of sight into account. With the line of sight, we form a tetrahedron shape instead of a triangle. As a result, the bispectrum with a line of sight now becomes sensitive to parity transformation. The bispectrum doesn't have a parity odd part when we have a density field which consists of purely RSD terms. For a density field, which also includes the relativistic RSD, such that: 
\begin{align} \label{parodd}
&\delta_g(\boldsymbol{k}) = \delta_{g\mathrm{N}} + \delta_{g\mathrm{GR}} \, , \\
&\delta_{g\mathrm{GR}} = \mathcal{K}(\boldsymbol{k})\delta(\boldsymbol{k})(\boldsymbol{n} \cdot \hat{\boldsymbol{k}}) 
\end{align}
where $\boldsymbol{n}$ is the line of sight and $\mathcal{K}(\boldsymbol{k})$ is the respective kernel in Fourier space. 
When $\boldsymbol{k}$ goes to $-\boldsymbol{k}$, which is the case for a parity transformation, (\ref{parodd}) is not invariant and results in an imaginary part to the bispectrum. In the case of the power spectrum, the imaginary parts combine to form a real contribution. As a result, the bispectrum can be broken down into a parity even and odd part. 
For a single tracer, it serves as the lowest order statistic, from which we can isolate the relativistic effects. In order to work out the even and odd parity bispectrum, we need the density contrast up to second order. The odd parity part of the bipsectrum has been worked out by \cite{Clarkson:2018dwn}, where they perform a spherical harmonic decomposition to isolate the parity-odd part and refer to it as the dipole. \cite{Jeong_2020} also works out the parity-odd part of the galaxy bispectrum and it has also been explicitly stated in \cite{Paul_2024}.
We can now define the parity odd part as 
\bea
B_{\mathrm{odd}} = \frac{1}{2}\Big[ B_g(\boldsymbol{k}_1, \boldsymbol{k}_2, \boldsymbol{k}_3) - B_g(-\boldsymbol{k}_1, -\boldsymbol{k}_2, -\boldsymbol{k}_3)\Big].
\eea
For the trispectrum, we investigate the same effects that arise from (\ref{parodd}).  We follow the same process as above to split it into odd and even parts.

\section{The relativistic trispectrum} \label{reltrispec}

We begin by calculating the full trispectrum, which is given by the Fourier counterpart of the 4-point function [\citealp{Verde_2001, Gualdi_2021}]. It corresponds to a quadrilateral configuration which is defined by four $k$ vectors. We can define $4$ density fields for the corresponding wave vectors and construct the trispectrum by cross-correlating $4$ density fields.  
Hence, in Fourier space, the observed galaxy trispectrum $T_{g}$ at fixed redshift should be given by:  
\begin{equation} \label{fulltrispec}
    \langle \Delta_g(\boldsymbol{k}_1) \Delta_g(\boldsymbol{k}_2) \Delta_g(\boldsymbol{k}_3) \Delta_g(\boldsymbol{k}_4) \rangle = (2\pi)^3 T_{g}(\boldsymbol{k}_1,\boldsymbol{k}_2,\boldsymbol{k}_3) \delta^D(\boldsymbol{k}_1+\boldsymbol{k}_2+\boldsymbol{k}_3+\boldsymbol{k}_4).
\end{equation}
Our convention for the matter density contrast is 
\bea 
\Delta = \Delta + \frac{1}{2} \Delta^{(2)} + \frac{1}{3!} \Delta^{(3)}.
\eea
The non-vanishing contributions in the connected trispectrum at tree level are:
\begin{equation} \label{trispeccoeffs}
 \begin{split}
    24\big\langle \Delta_g(\boldsymbol{k}_1) \Delta_g(\boldsymbol{k}_2) \Delta_g(\boldsymbol{k}_3) \Delta_g(\boldsymbol{k}_4)\big\rangle &= 4(\langle \Delta_g^{(1)}(\boldsymbol{k}_1) \Delta_g^{(1)}(\boldsymbol{k}_2) \Delta_g^{(1)}(\boldsymbol{k}_3) \Delta_g^{(3)}(\boldsymbol{k}_4)\rangle  + 3 \,\mathrm{cyc.} \,\mathrm{perms.} \\
    &+6( \langle \Delta_g^{(1)}(\boldsymbol{k}_1) \Delta_g^{(1)}(\boldsymbol{k}_2) \Delta_g^{(2)}(\boldsymbol{k}_3) \Delta_g^{(2)}(\boldsymbol{k}_4)\rangle + 5 \mathrm{\,cyc.} \mathrm{\,perms.} 
 \end{split}
\end{equation}
The factors of 4 and 6 come from the perturbative expansion of $\Delta_g$. Here, $\mathrm{cyc. \, perms.}$ stands for cyclic permutations and the permutations are in $\boldsymbol{k}$. In the first line of (\ref{trispeccoeffs}), we have the first and third-order contributions, while in the second line, we have the first and second-order contributions. In the first line, $\Delta$ is symmetric in $\boldsymbol{k}_1$, $\boldsymbol{k}_2$ and $\boldsymbol{k}_3$. Hence, the permutations are in $\boldsymbol{k}_4$ and result in $3$ other permutations. In the second line of (\ref{trispeccoeffs}), we have $\Delta$ symmetric in $\boldsymbol{k}_1$ and $\boldsymbol{k}_2$ and symmetric in $\boldsymbol{k}_3$ and $\boldsymbol{k}_4$. This results in $5$ more permutations. 
In order to calculate (\ref{trispeccoeffs}), we will Fourier transform the density field. The Fourier convention that we use is: 
\begin{equation} \label{Fourconv}
\begin{split}
    f(\boldsymbol{x}) &= \int \frac{\mathrm{d^3} k}{(2 \pi)^3} e^{i \boldsymbol{k} \cdot \boldsymbol{x}} f(\boldsymbol{k}), \, \, \,\, f(\boldsymbol{k}) = \int \mathrm{d^3} x e^{-i \boldsymbol{k} \cdot \boldsymbol{x}} f(\boldsymbol{x}).
\end{split}
\end{equation}
The first, second and third order density kernels can be written out in full in terms of the Fourier space kernels $\mathcal{K}^{(n)}$ and Dirac delta. Below, we have given $\Delta_g^{(1,2,3)}$ in terms of kernels. These kernels are also given explicitly in sections (\ref{firstandsec}) and (\ref{thirdorder}). 
\begin{align}
    \Delta_g^{(1)} (\boldsymbol{k}_2) &= \int \frac{\mathrm{d}^3 k_1}{(2 \pi)^3} \mathcal{K}^{(1)}(\boldsymbol{k}_1) \delta^{(1)} (\boldsymbol{k}_1) (2 \pi)^3 \delta^D(\boldsymbol{k}_1 - \boldsymbol{k}_2) \, , \label{delta1}\\ 
    \Delta_g^{(2)} (\boldsymbol{k}_3) &= \int \frac{\mathrm{d}^3 k_1}{(2 \pi)^3} \frac{\mathrm{d}^3 k_2}{(2 \pi)^3}\mathcal{K}^{(2)}(\boldsymbol{k}_1,\boldsymbol{k}_2,\boldsymbol{k}_3) \delta^{(1)} (\boldsymbol{k}_1)\delta^{(1)} (\boldsymbol{k}_2) (2 \pi)^3 \delta^D(\boldsymbol{k}_1 + \boldsymbol{k}_2 - \boldsymbol{k}_3) \, ,\label{delta2}\\
    \Delta_g^{(3)} (\boldsymbol{k}_4) &= \int \frac{\mathrm{d}^3 k_1}{(2 \pi)^3} \frac{\mathrm{d}^3 k_2}{(2 \pi)^3} \frac{\mathrm{d}^3 k_3}{(2 \pi)^3} \mathcal{K}^{(3)}(\boldsymbol{k}_1,\boldsymbol{k}_2,\boldsymbol{k}_3, \boldsymbol{k}_4) \delta^{(1)} (\boldsymbol{k}_1)\delta^{(1)} (\boldsymbol{k}_2) \delta^{(1)} (\boldsymbol{k}_3) (2 \pi)^3 \label{delta3} \\
    & \times \delta^D(\boldsymbol{k}_1 + \boldsymbol{k}_2 + \boldsymbol{k}_3 - \boldsymbol{k}_4). \notag
\end{align}
We can substitute (\ref{delta1}) and (\ref{delta3}) into the first and third order contribution for (\ref{trispeccoeffs}). Using Wick's theorem, 
we end up with 
\begin{equation}
    \begin{split}
    &\langle \Delta_g^{(1)}(\boldsymbol{k}_1) \Delta_g^{(1)}(\boldsymbol{k}_2) \Delta_g^{(1)}(\boldsymbol{k}_3) \Delta_g^{(3)}(\boldsymbol{k}_4) \rangle =  6 (2\pi)^3 \mathcal{K}^{(3)}(-\boldsymbol{k}_1, -\boldsymbol{k}_2, -\boldsymbol{k}_3, \boldsymbol{k}_4) \mathcal{K}^{(1)}(\boldsymbol{k}_1)\mathcal{K}^{(1)}(\boldsymbol{k}_2) \\
    & ~~~~~~~~~~~\times \mathcal{K}^{(1)}(\boldsymbol{k}_3) P(k_1) P(k_2) P(k_3) \delta^D(\boldsymbol{k}_1 + \boldsymbol{k}_2 + \boldsymbol{k}_3 + \boldsymbol{k}_4) + 3 \,\mathrm{cyc.} \, \mathrm{perms.}
    \end{split}
\end{equation}
and 
\begin{equation}
    \begin{split}
    \langle \Delta_g^{(1)}(\boldsymbol{k}_1) \Delta_g^{(1)}(\boldsymbol{k}_2) \Delta_g^{(2)}(\boldsymbol{k}_3) \Delta_g^{(2)}(\boldsymbol{k}_4) \rangle &=  4 (2 \pi)^3\Big[\mathcal{K}^{(1)}(\boldsymbol{k}_1) \mathcal{K}^{(1)}(\boldsymbol{k}_2) \mathcal{K}^{(2)}(-\boldsymbol{k}_1, \boldsymbol{k}_1+\boldsymbol{k}_3, \boldsymbol{k}_3) \\
    & \times \mathcal{K}^{(2)}(-\boldsymbol{k}_2, -\boldsymbol{k}_1-\boldsymbol{k}_3, \boldsymbol{k}_4) P(k_1) P(k_2) P(|\boldsymbol{k}_1 + \boldsymbol{k}_3|) \\
    &+ \mathcal{K}^{(1)}(\boldsymbol{k}_1) \mathcal{K}^{(1)}(\boldsymbol{k}_2) \mathcal{K}^{(2)}(-\boldsymbol{k}_2, \boldsymbol{k}_2+\boldsymbol{k}_3, \boldsymbol{k}_3)  \\
    & \times \mathcal{K}^{(2)}(-\boldsymbol{k}_1, -\boldsymbol{k}_2-\boldsymbol{k}_3, \boldsymbol{k}_4) P(k_1) P(k_2) P(|\boldsymbol{k}_2 + \boldsymbol{k}_3|)\Big] \\
    &\times \delta^D(\boldsymbol{k}_1 + \boldsymbol{k}_2 + \boldsymbol{k}_3 + \boldsymbol{k}_4) .
    \end{split}
\end{equation}
Using (\ref{fulltrispec}), we can now write the trispectrum as - 
\begin{equation} \label{trispeckernels}
  \begin{split}
&T_g(\boldsymbol{k}_1,\boldsymbol{k}_2,\boldsymbol{k}_3,\boldsymbol{k}_4) = 6 \mathcal{K}^{(3)}(-\boldsymbol{k}_1, -\boldsymbol{k}_2, -\boldsymbol{k}_3, \boldsymbol{k}_4) \mathcal{K}^{(1)}(\boldsymbol{k}_1)\mathcal{K}^{(1)}(\boldsymbol{k}_2) \mathcal{K}^{(1)}(\boldsymbol{k}_3) P(k_1) P(k_2) P(k_3) \\
&+ 3 \,\mathrm{cyc.} \,\mathrm{perms.} \\
    &+4 \Big[\mathcal{K}^{(1)}(\boldsymbol{k}_1) \mathcal{K}^{(1)}(\boldsymbol{k}_2) \mathcal{K}^{(2)}(-\boldsymbol{k}_1, \boldsymbol{k}_1+\boldsymbol{k}_3, \boldsymbol{k}_3)  \mathcal{K}^{(2)}(-\boldsymbol{k}_2, -\boldsymbol{k}_1-\boldsymbol{k}_3, \boldsymbol{k}_4) P(k_1) P(k_2) P(|\boldsymbol{k}_1 + \boldsymbol{k}_3|) \\
    &+ \mathcal{K}^{(1)}(\boldsymbol{k}_1) \mathcal{K}^{(1)}(\boldsymbol{k}_2) \mathcal{K}^{(2)}(-\boldsymbol{k}_2, \boldsymbol{k}_2+\boldsymbol{k}_3, \boldsymbol{k}_3)  \mathcal{K}^{(2)}(-\boldsymbol{k}_1, -\boldsymbol{k}_2-\boldsymbol{k}_3, \boldsymbol{k}_4) P(k_1) P(k_2) P(|\boldsymbol{k}_2 + \boldsymbol{k}_3|) \Big] \\
    &+ 5 \,\mathrm{cyc.} \,\mathrm{perm.}. \\
    \end{split}
\end{equation}

In order to compute the trispectrum, we require the first, second, and third-order Newtonian and relativistic kernels in Fourier space. The following section outlines the linear and second-order kernels. While the Newtonian third-order number counts have been derived previously, in this paper we will provide a detailed derivation of both the third order Newtonian and relativistic kernel in Fourier space.


\section{Third order galaxy number contrast in Fourier space} \label{thirdorderoverden}

\subsection{General expression for the number counts in general relativity}

In a galaxy survey, we can measure the fluctuations of galaxy number counts $\Delta$ as a function of the observed redshift and the photon direction $\boldsymbol{n}$. We define the galaxy number counts as 
\bea
    \Delta(z,\boldsymbol{ n} ) = \frac{N(z,\boldsymbol{ n}) - \langle N \rangle (z)}{\langle N \rangle (z)}
\eea
where $\langle .. \rangle$ is the average over direction $\boldsymbol{n}$ at fixed observed redshift $z$. The number of galaxies $N(z,\boldsymbol{n})$ can be written in terms of galaxy number density $\rho$, and $V$, which is the volume of the pixels of the galaxy map. 
\bea
   N(z,\boldsymbol{n}) = \rho(z,\boldsymbol{{n}}) \cdot V(z,\boldsymbol{{n}})
\eea
Performing Taylor expansions, we can obtain
\bea \label{equa2.3}
    \Delta(z,\boldsymbol{{n}}) = \delta(z,\boldsymbol{{n}}) - 3 \frac{\delta z}{1 + \Bar{z}} + \frac{\delta V(z,\boldsymbol{{n}})}{\Bar{V}}.
\eea 
This shows us that the number counts are not only affected by the underlying dark matter distribution, and fluctuations which occur in the observed redshift, but also fluctuations which are observed in the volume pixel of the galaxy maps. These fluctuations are distortions, which are induced by the peculiar velocities of the galaxies observed in the volume of the pixel. When we are inside the horizon, the terms simplify to give rise to the redshift space distortions. Kaiser \cite{Kaiser:1987qv} showed that going from real space to redshift space, we have  
\bea \label{equa 2.4}
   \Delta(z,\boldsymbol{{n}}) = \delta(z,\boldsymbol{{n}}) - \frac{1}{\mathcal{H}} \partial_r v_{||} - \frac{1}{\mathcal{H} r} \Big( 2 + \frac{\mathrm{d} \mathrm{ln} \phi}{\mathrm{d} \mathrm{ln} r}  \Big) v_{||}
\eea 
 where  $v_{||} = \boldsymbol{v}\cdot \boldsymbol{{n}}$ is the peculiar velocity along the line of sight, $r$ is the comoving radial  distance, $\mathcal{H}$ is the comoving Hubble parameter and $\phi$ is the selection function. Using the Poisson and Euler equations, we can relate the density fluctuation and the peculiar velocity to the metric potential, where $ \delta \sim (k/\mathcal{H})^2 \phi$ and $v \sim (k/\mathcal{H}) \psi$. This tells us that the first two terms in (\ref{equa 2.4}) scale as $(\mathcal{H}/k)^0$, while the last term scales by $(\mathcal{H}/k)^1$. For the modes which are inside the Hubble radius, the first two terms dominate and the last term is negligible. However, when we go to large scales, where $k \sim \mathcal{H}$, the last term also starts to dominate and cannot be neglected. 

However, (\ref{equa 2.4}) is not representative of the full picture. The derivation follows a Newtonian approach and doesn't contain all of the redshift and the volume contributions. To be consistent and include all the distortions which are contributing to $\Delta$, we should perform a fully relativistic calculation. From (\ref{equa 2.4}), both the redshift and volume perturbations are calculated to include all the distortions. The redshift perturbation can be written:
\begin{align} \label{redfluc}
    \delta(z,\boldsymbol{{n}}) = b\delta - 3 v_{||} + 3 \Psi + 3 \int^r_0 d\tilde{r}\,({\Phi}' + {\Psi}')
\end{align}
These fluctuations are calculated by solving the null geodesic equation and evaluate the variation in evergy between the emission and the observation. 
The relativistic volume perturbation can be written as: 
\begin{align} \label{volfluc}
    \frac{\delta V}{\bar{V}}(z,\boldsymbol{n}) &= -2(\Psi + \Phi) + 4 v_{||} + \frac{1}{\mathcal{H}} \Big[ {\Phi}' + \partial_r\Psi - \frac{d v_{||} }{d r}\Big] \\ \notag
    &+\Big( \frac{{\mathcal{H}'}}{\mathcal{H}^2} + \frac{2}{r\mathcal{H}}\Big) \Big[\Psi - v_{||} + \int^r_0 d \tilde{r}\, ({\Phi}' + {\Psi}')\Big] - 3\int^r_0 d \tilde{r}\,({\Phi}' + \Psi') \\ \notag
    &+\frac{2}{r} \int^r_0 d \tilde{r}\, (\Phi + \Psi) - \frac{1}{r} \int^r_0 d \tilde{r}\frac{r-\tilde{r}}{\tilde{r}} \Delta_{\Omega}(\Phi + \Psi)
\end{align}
where $\Delta_{\Omega}$ is the laplacian transverse to the line-of-sight. To calculate the volume fluctuations, we first map the physical volume of a pixel to its observed volume, accounting for the perturbations in the photon's direction and the length of the geodesic. These perturbations contribute to the fluctuations in the observed volume. The explicit calculation involves solving the null geodesic equation in a perturbed Friedmann universe. When combining (\ref{redfluc}) and (\ref{volfluc}), we arrive to the linear relativistic galaxy number counts, which have been computed in [ \citealp{Bonvin_2011}, \citealp{Dio_2019}\footnote{Our $\boldsymbol{n}$ is minus theirs;  their convention $\delta_g^{(1)}  + \delta_g^{(2)}+ \delta_g^{(3)}$ differs from ours: $\delta_g^{(1)}  + \delta_g^{(2)}/2+ \delta_g^{(3)}/3!$. We alter the standard kernels $F_2, G_2, F_3, G_3$ accordingly.}, \citealp{Yoo_2009}]. The expression is as follows:
\begin{equation} \label{fullnumc}
    \begin{split}
        \Delta(z,\boldsymbol{{n}}) &= b\delta - \frac{1}{\mathcal{H}} \partial_r v_{||} \\
        &- \Big(\frac{{\mathcal{H}'}}{\mathcal{H}^2} + \frac{2 - 5s}{r \mathcal{H}} + 5s -b_{\rm e} \Big) v_{||} \\
        &  + (5s - 2) \int ^{r}_{0} \frac{r - \tilde{r}}{2r\tilde{r}} \Delta_{\Omega} (\Phi + \Psi) d\tilde{r}\\
        & +(b_{\rm e} - 3) \mathcal{H}V + (5s - 2)\Phi + \Psi + \frac{1}{\mathcal{H}}  {\Phi}' + \frac{2 - 5s}{r} \int^{r}_{0} d\tilde{r}\,(\Phi + \Psi) \\
        &+ \Big(\frac{{\mathcal{H}'}}{\mathcal{H}^2} + \frac{2 + 5s}{r \mathcal{H}} + 5s -b_{\rm e} \Big) \Big(\Psi + \int^{r}_{0} d\tilde{r}\,(\Phi' + {\Psi}' ) \Big) .
    \end{split}
\end{equation}
The dash refers to a partial conformal time derivative and the integrals are along the path along the past light cone. The underlying dark matter distribution is correlated with the galaxies using three bias parameters. Here, $b$ is the galaxy bias, $b_{\rm e}=\partial {\ln} (a^3 \bar{n}_g)/\partial \ln a$ is the evolution and $s=-(2/5)\partial \ln \bar{n}_g/\partial \ln L$ is the magnification bias. 


The first term in (\ref{fullnumc}) refers to the true fluctuations in the distribution of galaxies. We are especially interested in the terms that are proportional to $v_{||}$ and scale $\mathcal{H}/k$. They arise from the peculiar velocities of the galaxies and are represented by the terms in the second line of (\ref{fullnumc}). From (\ref{redfluc}) and (\ref{volfluc}), we can see that they are sourced both by the redshift and volume perturbations. These effects, unlike the standard redshift-space distortions terms, which are represented by the terms in the first line of (\ref{fullnumc}), are not symmetric along the line of sight. They are different in nature than the standard RSD terms, as there are no radial derivatives. They are also responsible for generating the odd multipoles in $n$-point spectra. When observing spheres in real space, these effects distort them into egg shapes in redshift space. This is in contrast to the standard RSD, which distorts spheres into ellipses instead. 
\cite{Bonvin_2014} gave a schematic representation of this effect for two galaxies. When galaxies are observed on a light cone at separate times, the positions of the galaxies change due to their peculiar velocity.  As a result, the real space line-of-sight separation is not the same as redshift space line-of-sight separation. 

In this paper, we neglect the terms that are integrated along the line of sight. These include lensing magnification and integrated Sachs-Wolfe effects. We also neglect the terms that involve gravitational potentials, which are responsible for the effect arising from gravitational redshift and are also asymmetric along your sight, although we leave it for a future analysis. 

\subsection{First and second order} \label{firstandsec}
 
We define the Newtonian and relativistic number counts as 
\bea
    \mathcal{K}^{(n)} = \mathcal{K}^{(n)}_\mathrm{N} +\mathcal{K}^{(n)}_\mathrm{R}
\eea
where $n = 1,2,3$. The linear order kernels in Fourier space have been worked out before. \cite{Yoo_2009, Yoo_2010, Bonvin_2011,Challinor:2011bk, McDonald_2009}. In Fourier space, we can write down the linear order kernel as 
\begin{equation} \label{newt1}
    \mathcal{K}^{(1)}_{\mathrm{N}}(\boldsymbol{k}) = b_1 + f\mu^2 \, ,
\end{equation}
\begin{equation} \label{gr1}
    \mathcal{K}^{(1)}_{\mathrm{GR}}(\boldsymbol{k}) = i\frac{\mu}{\mathcal{H}} f\Big[b_e -2s - \frac{2(1-s)}{r\mathcal{H}} - \frac{ \mathcal{H}'}{\mathcal{H}}\Big]
\end{equation}
where $\mu_i = \hat{\boldsymbol{k}}_i\cdot\boldsymbol{n}$.  We have neglected the terms which scale $(\mathcal{H}/k)^2$ and above. $\mathcal{K}^{(1)}_{\mathrm{N}}(\boldsymbol{k})$ is the standard RSD effect taking into account the Kaiser approximation. $\mathcal{K}^{(1)}_{\mathrm{GR}}(\boldsymbol{k})$ is used to describe the Doppler effect. If we closely examine (\ref{newt1}), we can see that for the Newtonian kernel, we only have even powers of $\mu$. On the other hand, in (\ref{gr1}), we have odd powers of $\mu$. So, $\mathcal{K}_{\mathrm{N}}^{(1)}$ is still invariant when $\boldsymbol{k} \rightarrow -\boldsymbol{k}$, however $\mathcal{K}_{\mathrm{GR}}^{(1)}$ is not.  This effect is going to leave an imprint on the correlation functions, mainly when $k \approx \mathcal{H}$. 

In order to calculate the second order Newtonian and relativistic kernels, we require the matter density contrast and the velocity and the metric potentials at second order. They are given by:
\begin{equation}
\begin{split}
   \delta^{(2)}(\boldsymbol{k}) = \int \frac{\mathrm{d^3} k_1}{(2 \pi)^3} \frac{\mathrm{d^3} k_2}{(2 \pi)^3} \delta^{(1)}(\boldsymbol{k}_1) \delta^{(1)}(\boldsymbol{k}_2) F_2(\boldsymbol{k}_1, \boldsymbol{k}_2) (2\pi)^3 \delta^D(\boldsymbol{k}_1 + \boldsymbol{k}_2 - \boldsymbol{k}) \, ,
\end{split}
\end{equation}
\begin{equation}
\begin{split}
   v^{(2)}(\boldsymbol{k}) = f \frac{\mathcal{H}}{k^2}\int \frac{\mathrm{d^3} k_1}{(2 \pi)^3} \frac{\mathrm{d^3} k_2}{(2 \pi)^3} \delta^{(1)}(\boldsymbol{k}_1) \delta^{(1)}(\boldsymbol{k}_2) G_2(\boldsymbol{k}_1, \boldsymbol{k}_2) (2\pi)^3 \delta^D(\boldsymbol{k}_1 + \boldsymbol{k}_2- \boldsymbol{k}) \, ,
\end{split}
\end{equation}
\begin{equation}
    \Phi^{(2)} (\boldsymbol{k}) = \Psi^{(2)} (\boldsymbol{k}) = - \frac{3}{2} \Omega_m \frac{\mathcal{H}^2}{k^2} \delta^{(2)}(\boldsymbol{k}) .
\end{equation}
The second order number counts have been derived independently by several groups \cite{Yoo:2014kpa, Bertacca:2014dra, Dio_2014} and are given below:
\bea
\label{Delta2N}
\Delta^{(2)}_N &=&\delta_g^{(2)} - \mathcal{H}^{-1} \partial_r v_{||}^{(2)} 
- \mathcal{H}^{-1} \partial_r \left( v_{||}  \delta_g \right)  + \mathcal{H}^{-2}  \partial_r \left( v_{||} \partial_r v_{||} \right) \, ,
\\
\label{Delta2R}
\Delta^{(2)}_R &=-&\left( \frac{{\mathcal{H}'}}{\mathcal{H}^2} + \frac{2 - 5s}{\mathcal{H} r}+ 5s - b_e\right) v_{||}^{(2)}
+ \left( 1 + 3 \frac{{\mathcal{H}'}}{\mathcal{H}^2} + \frac{4 - 5s}{\mathcal{H} r} + 5s - 2 b_e \right) \mathcal{H}^{-1} v_{||} \partial_r v_{||}
\nonumber \\
&&
 - \left( \frac{{\mathcal{H}'}}{\mathcal{H}^2} + \frac{2 - 5s}{\mathcal{H} r} + 5s - b_e \right) v_{||} \delta_g  
+ \mathcal{H}^{-1} v_{||}  \delta_g' - 2 \mathcal{H}^{-1} v^a_{\perp} \partial_{a{\perp}} v_{||}
\nonumber \\
&&
 -  \mathcal{H}^{-2} \psi \partial_r^2 v_{||} + \mathcal{H}^{-1} \psi \partial_r \delta_g + \mathcal{H}^{-2} v_{||} \partial_r^2 \psi \, .
\eea
Following that, the Fourier kernels have also been worked out by \cite{Bertacca:2014hwa} and \cite{Jeong_2020}. The Newtonian and relativistic part of the kernels are given explicitly below: 
\begin{equation}
\begin{split}
    \mathcal{K}^{(2)}_{\mathrm{N}}(\boldsymbol{k}_1, \boldsymbol{k}_2) &= b_1 F_2(\boldsymbol{k}_1, \boldsymbol{k}_2) + b_2 + fG_2(\boldsymbol{k}_1, \boldsymbol{k}_2) \mu_3^2 \\
    & + f^2\frac{\mu_1\mu_2}{k_1k_2}(\mu_1 k_1 + \mu_2 k_2)^2 + b_1 \frac{f}{k_1k_2} \Big[ (\mu_1^2 + \mu_2^2) + \mu_1\mu_2(k_1^2 + k_2^2)\Big] \, ,
    \end{split}
\end{equation}

\begin{equation}
\begin{split}
    \mathcal{K}^{(2)}_{\mathrm{GR}}(\boldsymbol{k}_1, \boldsymbol{k}_2) &= i \frac{\mathcal{H}}{k_1^2k_2^2} \Big[-\frac{3}{2} \Omega_m b_1(\mu_1k_1^3 + \mu_2k_2^3) + 2f^2 (\mu_1k_1 + \mu_2k_2)(\boldsymbol{k}_1 \cdot \boldsymbol{k}_2) \\ 
    &+ f \Big\{b_1 \Big(f + b_e - 2s - \frac{2(1-s)}{r\mathcal{H}} - \frac{\mathcal{H}'}{\mathcal{H}^2} \Big) + \frac{b_1'}{\mathcal{H}}\Big\} k_1k_2 (\mu_1k_1 + \mu_2k_2) \\ 
    &- \frac{3}{2}\Omega_m f (\mu_1^3 k_1^3 + \mu_2^3 k_2^3 ) \\ 
    &+ \Big\{ \frac{3}{2}\Omega_mf - f^2\Big(3 - 2b_e + 4s + \frac{4(1-s)}{r\mathcal{H}} + \frac{3\mathcal{H}'}{\mathcal{H}}\Big) \Big\}\mu_1 \mu_2k_1k_2(\mu_1 k_2 + \mu_2 k_1) \\ 
    &+ f \Big(b_e - 2s - \frac{2(1-s)}{r\mathcal{H}} - \frac{\mathcal{H}'}{\mathcal{H}^2} \Big)\frac{\mu_3 k_1^2k_2^2}{k_3}G_2(\boldsymbol{k}_1, \boldsymbol{k}_2) \Big] .
\end{split}
\end{equation}
We will use $\mathcal{K}_{\mathrm{N}}^{(2)}$ and $\mathcal{K}_{\mathrm{GR}}^{(2)}$ to calculate the second order kernels in  (\ref{trispeckernels}). 

\subsection{Third Order} \label{thirdorder}

For the trispectrum, we will need to go up to third order in perturbation theory. Following from section \ref{firstandsec}, we will now require the matter density contrast, the velocity and metric potentials at third order. They are given below:  
\begin{equation}
\begin{split}
   \delta^{(3)}(\boldsymbol{k}) =& \int \frac{\mathrm{d^3} k_1}{(2 \pi)^3} \frac{\mathrm{d^3} k_2}{(2 \pi)^3} \frac{\mathrm{d^3} k_3}{(2 \pi)^3} \delta^{(1)}(\boldsymbol{k}_1) \delta^{(1)}(\boldsymbol{k}_2) \delta^{(1)}(\boldsymbol{k}_3) F_3(\boldsymbol{k}_1, \boldsymbol{k}_2, \boldsymbol{k}_3) (2\pi)^3 \\
   & \times \delta^D(\boldsymbol{k}_1 + \boldsymbol{k}_2 +\boldsymbol{k}_3- \boldsymbol{k}) \, ,
\end{split}
\end{equation}
\begin{equation}
\begin{split}
   v^{(3)}(\boldsymbol{k}) &= f \frac{\mathcal{H}}{k^2}\int \frac{\mathrm{d^3} k_1}{(2 \pi)^3} \frac{\mathrm{d^3} k_2}{(2 \pi)^3} \frac{\mathrm{d^3} k_3}{(2 \pi)^3} \delta^{(1)}(\boldsymbol{k}_1) \delta^{(1)}(\boldsymbol{k}_2) \delta^{(1)}(\boldsymbol{k}_3) G_3(\boldsymbol{k}_1, \boldsymbol{k}_2, \boldsymbol{k}_3) (2\pi)^3 \\
   & \times \delta^D(\boldsymbol{k}_1 + \boldsymbol{k}_2 +\boldsymbol{k}_3- \boldsymbol{k}) \, ,
\end{split}
\end{equation}
\begin{equation}
    \Phi^{(3)} (\boldsymbol{k}) = \Psi^{(3)} (\boldsymbol{k}) = - \frac{3}{2} \Omega_m \frac{\mathcal{H}^2}{k^2} \delta^{(3)}(\boldsymbol{k}) .
\end{equation}
A consistent galaxy bias expansion to third order is also introduced 
\bea \label{conv}
    \delta_g^{(3)} = b_1 \delta^{(3)} + 3b_2 \delta \delta^2 + b_3 \delta^3 .
\eea   
\cite{Dio_2019} have computed the number counts to third order by taking into account all the relativistic effects that scale $\mathcal{H}/k$. From (\ref{equa2.3}), the density, redshift and volume perturbations are all extended beyond linear theory. The terms that contribute to the third order number counts are:
\bea
\Delta^{(3)}(z, \boldsymbol{n}) = \delta^{(3)}(z,\boldsymbol{n}) + 3\delta^{(2)}(z,\boldsymbol{n}) \frac{\delta V(z, \boldsymbol{n})}{\bar{V}(z)} + 3\delta(z,\boldsymbol{n}) \frac{\delta V^{(2)}(z, \boldsymbol{n})}{\bar{V}(z)} + \frac{\delta V^{(3)}(z, \boldsymbol{n})}{\bar{V}(z)}. 
\eea
The third order redshift and volume contributions can be worked out by extending the peculiar velocity to third order. Combining both the perturbations, below presented are the Newtonian and the relativistic third order number counts. 
\bea
\label{Delta3N}
\Delta_N^{(3)} &=& { \delta_g^{(3)} -\frac{{\partial_r v_{||}^{(3)}}}{\mathcal{H}}}
{ 
-\left[ \mathcal{H}^{-1} \partial_r\left(  v_{||}\delta_g  \right)\right]^{(3)}
+ \left[ \mathcal{H}^{-2}  \partial_r \left( v_{||} \partial_r v_{||} \right) \right]^{(3)}
}
\nonumber \\
&&
 - { \frac{1}{6} \mathcal{H}^{-3}\partial_r^3 v_{||}^3
 +\frac{1}{2} \mathcal{H}^{-2} \partial_r^2 \left( \delta_g v_{||}^2 \right)
 } \, ,
 \\
 \Delta_R^{(3)} &=&
 \left\{
{-\left( v_{||}^{(3)} + \left[v_{||} \delta_g\right]^{(3)}  \right)\left(\frac{{{\mathcal{H}'}}}{{{\mathcal{H}}}^2}+\frac{2- 5s}{{{\mathcal{H}}} r} +5s - b_e \right)} 
\right.
\nonumber \\
&&
+ \left( 1 + 3 \frac{{\mathcal{H}'}}{{\mathcal{H}}^2} + \frac{4- 5s}{{\mathcal{H}} r} +5s - 2 b_e \right) {\mathcal{H}}^{-1} \left[ v_{||} \partial_r v_{||} \right]^{(3)}
 + {\mathcal{H}}^{-1} \left[ v_{||}  \delta_g' \right]^{(3)}
\nonumber\\
&&
\left.
 - 2 {\mathcal{H}}^{-1} \left[ v_a \partial^a v_{||} \right]^{(3)}
  -  {\mathcal{H}}^{-2} \left[ \psi \partial_r^2 v_{||} \right]^{(3)}
 + {\mathcal{H}}^{-1} \left[ \psi \partial_r \delta_g \right]^{(3)} + {\mathcal{H}}^{-2} \left[ v_{||} \partial_r^2 \psi \right]^{(3)}
 \right\}
 \nonumber \\
 &&
 +
 {
 3 {\mathcal{H}}^{-3} \partial_r v_{||} \left( \partial_r^2 v_{||} \psi - v_{||} \partial_r^2 \psi \right)
 + {\mathcal{H}}^{-3} v_{||} \left(  \partial_r^3 v_{||} \psi - v_{||} \partial_r^3 \psi \right)
 }
   \nonumber \\
 &&
  - {\mathcal{H}}^{-2} \psi \partial_r^2 \left( v_{||} \delta_g \right) 
 + {\mathcal{H}}^{-2} \partial_r^2\psi  \left( v_{||} \delta_g \right) 
 -
 \frac{1}{3{\mathcal{H}}^2} \partial^2_r \left( v_{||}^3 \right) \left( 1 + \frac{3  {\mathcal{H}'}}{{\mathcal{H}}^2} +\frac{3}{{\mathcal{H}} r} - \frac{3}{2} b_e \right)
      \nonumber \\
 &&
 +\frac{1}{2 {\mathcal{H}}} \partial_r \left( v_{||}^2 \delta \right)  \left( 1 + 3 \frac{{\mathcal{H}'}}{{\mathcal{H}}^2} +\frac{4}{{\mathcal{H}} r}  - 2b_e\right)
 - \frac{1}{{\mathcal{H}}^2} \partial_r \left( v_{||}^2  \delta' \right)
    \nonumber \\
 &&
 + \frac{1}{2 {\mathcal{H}}^2}\partial_r^2 \left( v_{||} v_{\perp}^a v_{\perp a} \right) - \frac{1}{{\mathcal{H}}}\partial_r \left( \delta  v^a_{\perp} v_{\perp a} \right) 
 +\frac{1}{2{\mathcal{H}}} v^2\partial_r \delta
 + \frac{2}{{\mathcal{H}}^2} \partial_r \left( \partial_a v_{||} v^a_{\perp} v_{||} \right)  \, .
 \label{Delta3R}
\eea
Following the third order number counts, we can calculate the equivalent kernels in Fourier space. The Newtonian kernel in Fourier space has been calculated by \cite{Bernardeau_2002} and \cite{Philcox_2022}. We have independently derived the Newtonian and the relativistic kernels by Fourier transforming (\ref{Delta3N}) and (\ref{Delta3R}). As evident, there are a variety of coupling terms between different orders.  We also require the radial and transverse derivatives of the velocity potential. The details of the necessary equations and examples of calculated terms of different natures are given in Appendix \ref{Fourtrans}.

At third order, the Newtonian part of the kernel is:

\begin{equation} \label{newtkern}
    \begin{split}
    \mathcal{K}^{(3)}_N &= \Bigg\{b_1F_3(\boldsymbol{k}_1, \boldsymbol{k}_2, \boldsymbol{k}_3) + 3b_2F_2(\boldsymbol{k}_1, \boldsymbol{k}_2) + b_3 + \mu_4^2fG_3(\boldsymbol{k}_1, \boldsymbol{k}_2, \boldsymbol{k}_3) \\ 
    &+ \frac{3\mu_3}{k_3}(\mu_1 k_1 + \mu_2 k_2 + \mu_3 k_3 )f\Bigg[b_1F_2(\boldsymbol{k}_1, \boldsymbol{k}_2) + b_2 + \frac{(\mu_1 k_1 +\mu_2k_2)^2}{|\boldsymbol{k}_1 + \boldsymbol{k}_2|^2}fG_2(\boldsymbol{k}_1, \boldsymbol{k}_2) \Bigg] \\ 
    &+ \frac{3(\mu_1 k_1 +\mu_2k_2)}{|\boldsymbol{k}_1 + \boldsymbol{k}_2|^2}(\mu_1 k_1 +\mu_2k_2 + \mu_3 k_3) f G_2(\boldsymbol{k}_1, \boldsymbol{k}_2)(b_1 + \mu_3^2f)  \\ 
    &+ \frac{\mu_2 \mu_3}{ k_2 k_3}(\mu_1 k_1 + \mu_2 k_2 + \mu_3 k_3)^2 f^2 \Big[\frac{\mu_1}{k_1}(\mu_1 k_1 + \mu_2 k_2 + \mu_3 k_3)f + 3b_1 \Big]\Bigg\}_{\circlearrowright_{1\cdots3}} .
    \end{split}
\end{equation}
Here, $\circlearrowright$ denotes symmeterisation on $\boldsymbol{k}_{1..3}$. $F_3$ and $G_3$ denote the third order kernels in the Fourier space. Following the same idea from the second order kernels, the Newtonian kernel has all even powers of $\mu$. Hence, it is invariant under a parity transformation. We expect the parity odd signal to rise purely from the relativistic kernel. The relativistic part of the kernel follows from Fourier transforming the third order number counts in (\ref{Delta3R}): 

\begin{equation} \label{relkern}
\begin{split}
\mathcal{K}^{(3)}_{\mathrm{GR}} &=   \mathrm{i} f \mathcal{H} \Bigg \{ - \Bigg(\frac{{\mathcal{{H}'}}}{\mathcal{H}^2} + \frac{2 - 5s}{\mathcal{H}r} + 5s - b_e \Bigg) \Bigg[\frac{\mu_4}{k_4}  G_3(\boldsymbol{k}_1, \boldsymbol{k}_2, \boldsymbol{k}_3) +  \frac{3\mu_3}{k_3} \big[b_1 F_2(\boldsymbol{k}_1 , \boldsymbol{k}_2) +  b_2\big] 
\\
&+  \frac{3(\mu_1k_1 + \mu_2 k_2)}{|\boldsymbol{k}_1 + \boldsymbol{k}_2|^2} b_1G_2(\boldsymbol{k}_1, \boldsymbol{k}_2) \Bigg] -3f G_2(\boldsymbol{k}_1, \boldsymbol{k}_2) \Bigg(1 +  \frac{3{\mathcal{H}'}}{\mathcal{H}^2} + \frac{4-5s}{\mathcal{H}r} + 5s - 2 b_e \Bigg) \\
&\Bigg[ \frac{\mu_3}{k_3}\Bigg(\frac{(\mu_1k_1 + \mu_2 k_2)}{|\boldsymbol{k}_1+ \boldsymbol{k}_2|}\Bigg)^2 + \frac{\mu^2_3(\mu_1k_1 + \mu_2 k_2)}{|\boldsymbol{k}_1+ \boldsymbol{k}_2|^2} \Bigg]  + 3f \Bigg( \frac{\mu_3}{k_3} \Big[ b_1 F_2(\boldsymbol{k}_1 , \boldsymbol{k}_2) +  b_2\Big] \\
&+  \frac{(\mu_1k_1 + \mu_2 k_2)}{|\boldsymbol{k}_1 + \boldsymbol{k}_2|^2}  b_1 G_2(\boldsymbol{k}_1, \boldsymbol{k}_2)  \Bigg) \\
&+ 6 f  G_2(\boldsymbol{k}_1, \boldsymbol{k}_2) \Bigg[\frac{(\mu_1k_1 + \mu_2 k_2)  (\boldsymbol{k}_3 \cdot (\boldsymbol{k}_1 + \boldsymbol{k}_2) - \mu_3 k_3 (\mu_1k_1 + \mu_2 k_2))}{|\boldsymbol{k}_1 + \boldsymbol{k}_2|^2k_3^2}  \\  
&+ \frac{\mu_3\Big\{(\boldsymbol{k}_1 + \boldsymbol{k}_2) \cdot \boldsymbol{k}_3 - (\mu_1k_1 + \mu_2 k_2)\mu_3k_3\Big\}}{|\boldsymbol{k}_1 + \boldsymbol{k}_2|^2k_3}  \Bigg] 
\\
& - \frac{9}{2} \Omega_m \Bigg\{ \Bigg[ \frac{(\mu_1 k_1 + \mu_2 k_2)^3}{|\boldsymbol{k}_1 + \boldsymbol{k}_2|^2k_3^2} G_2(\boldsymbol{k_}1, \boldsymbol{k}_2) +  \frac{\mu_3^3 k_3}{|\boldsymbol{k}_1 + \boldsymbol{k}_2|^2} F_2(\boldsymbol{k}_1, \boldsymbol{k}_2) \Bigg]  \\
&+ \frac{1}{f} \Bigg\{ \frac{(\mu_1 k_1 + \mu_2 k_2)}{k_3^2} \Big[b_1 F_2(\boldsymbol{k}_1, \boldsymbol{k}_2) +  b_2\Big]  + \frac{\mu_3 k_3}{|\boldsymbol{k}_1 + \boldsymbol{k}_2|^2 }  b_1 F_2(\boldsymbol{k}_1, \boldsymbol{k}_2)\Bigg\}  \\
&-   \Bigg[ \frac{\mu_3^2(\mu_1 k_1 + \mu_2 k_2)}{|\boldsymbol{k}_1 + \boldsymbol{k}_2|^2}  G_2(\boldsymbol{k}_1, \boldsymbol{k}_2)   +  \frac{\mu_3 (\mu_1 k_1 + \mu_2 k_2)^2}{|\boldsymbol{k}_1 + \boldsymbol{k}_2|^2k_3}  F_2(\boldsymbol{k}_1, \boldsymbol{k}_2)  \Bigg] \Bigg\} \\
&- 6\Bigg(\frac{9}{2} \Omega_m f \mu_1^2 \mu_2 \Bigg(\frac{\mu_2^2 k_2}{k_3^2}  - \frac{\mu_3^2}{k_2}  \Bigg)  - \frac{3}{2} \Omega_m \Bigg\{f\frac{\mu_1}{k_1} \Bigg(\frac{ \mu_2^4 k_2^2}{ k_3^2}  - \frac{ \mu_2 \mu_3^3 k_3}{ k_2}   \Bigg) \\
&-  \frac{\mu_2}{k_2}   \Bigg[ \frac{(\mu_2 k_2 + \mu_3 k_3)^2}{k_1^2 }   - \mu_1^2  b_1  \Bigg] \Bigg\} \Bigg) \\
&+6\Bigg(- \Big(\mu_1 k_1 + \mu_2 k_2 + \mu_3 k_3\Big) \Bigg( \frac{\mu_1 \mu_2 \mu_3}{3k_1k_2k_3}f^2 \Bigg(1 + \frac{3{\mathcal{H}'}}{\mathcal{H}^2} + \frac{3}{\mathcal{H}r} - \frac{3}{2} b_e \Bigg)  \\
&\Big(\mu_1 k_1 + \mu_2 k_2 + \mu_3 k_3\Big)  
+ \frac{\mu_1 \mu_2}{2 k_1 k_2} f \Bigg(1 + \frac{3{\mathcal{H}'}}{\mathcal{H}^2} + \frac{4}{\mathcal{H}r} - 2 b_e \Bigg)   - \frac{\mu_1 \mu_2}{k_1 k_2}f^2 \Bigg) 
\\
&+ \big(\boldsymbol{k}_2 \cdot \boldsymbol{k}_3 - \mu_2 k_2 \mu_3 k_3\big)\big(\mu_1 k_1 + \mu_2 k_2 + \mu_3 k_3\big) \Bigg(\Bigg[  \frac{f^2\mu_1\big(\mu_1 k_1 + \mu_2 k_2 + \mu_3 k_3\big)}{2k_1 k_2^2 k_3^2}  +  \frac{f}{k_2^2 k_3^2} \Bigg]\Bigg)
\\
&  + 2f^2 \frac{ \mu_1 \mu_3}{k_1 k_2^2 k_3}   \Bigg[\big(\boldsymbol{k}_1 \cdot \boldsymbol{k}_2 - \mu_1 k_1 \mu_2 k_2 \big) \big(\mu_1 k_1 + \mu_2 k_2 + \mu_3 k_3\big)\Bigg] - \frac{f}{2}  \frac{ \mu_3 k_3}{k_1^2 k_2^2} (\boldsymbol{k}_2 \cdot \boldsymbol{k}_3) \Bigg) \Bigg\} _{_{\circlearrowright_{1\cdots3}}}.
\end{split}
\end{equation}
The full expression for the third-order relativistic terms is extremely long and we have given the terms that dominate the relativistic corrections. These terms are all odd powers of $\mu$. This suggests that the kernel will be sensitive to a parity transformation and we will get a signal for a parity-odd trispectrum. We reiterate, that as is the case in the power spectrum and the bispectrum, they are suppressed by $\mathcal{H}/k$ with respect to the Newtonian kernel. From the above expression, we can see that the relativistic kernel is dependent on the magnification and the evolution biases. This makes it sensitive to the different surveys. In section \ref{numrel}, we have shown numerical results that investigate this contribution for different surveys such as Euclid and SKAO. 

\section{Numerical results} \label{numrel}
We compute numerically the relativistic trispectrum that is introduced in section~\ref{reltrispec}. We have defined the even and odd parity trispectrum and the geometry used to evaluate it. The matter power spectrum has been computed using CAMB \cite{Lewis_2000}.  

\subsection{Geometry for the trispectrum}

 \begin{figure} [h]
    \centering
    \includegraphics[width=0.89\linewidth]{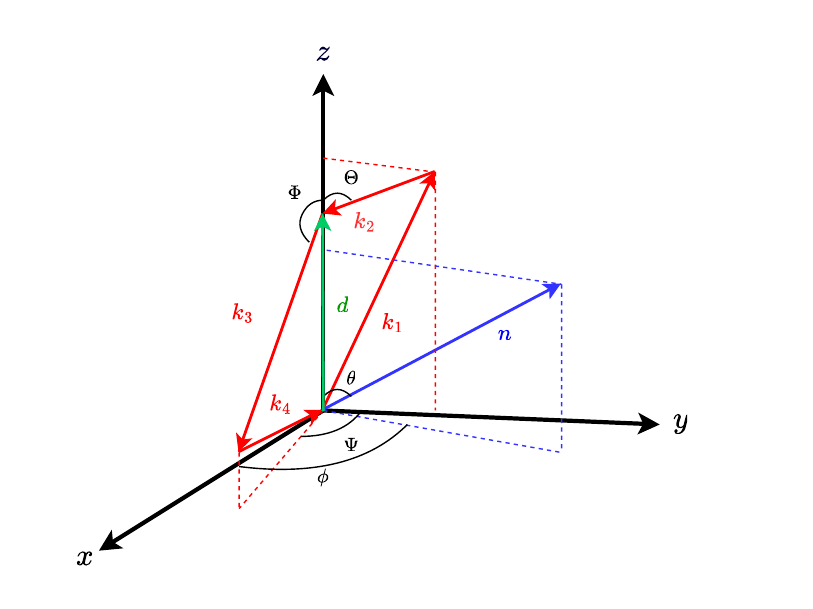}
    \caption{Overview of the relevant vectors and angles for the Fourier space trispectrum.}
    \label{fig:IMgeom}
\end{figure}
The co-ordinate basis for the vectors in Fig.~\ref{fig:IMgeom} are 
\begin{align}
    \boldsymbol{d} &= (0,0,d) \,, \\ 
    \boldsymbol{n} &= (\sin \theta\cos\phi,\sin\theta\sin\phi,\cos\theta) \,, \\ 
    \boldsymbol{k}_1 &= d(0, s \sin\Theta, 1+s\cos\Theta ) \,,\\ 
    \boldsymbol{k}_2 &= -d(0,s \sin\Theta, s \cos\Theta) \,, \\ 
    \boldsymbol{k}_3 &= d(t \sin\Phi \sin\Psi, t \sin\Phi \cos\Psi,  t\cos\Phi ) \,, \\ 
    \boldsymbol{k}_4 &= -d(t \sin\Phi \sin\Psi, t \sin\Phi \cos\Psi, t \cos\Phi + 1) .
\end{align}
The trispectrum can now be expressed in terms of $k_1$, $k_2$, $k_3$, $\theta$, $\phi$, $\Theta$, $\Phi$ and $\Psi$. The vector, $\boldsymbol{k} = \boldsymbol{k}_1 + \boldsymbol{k}_2 = - \boldsymbol{k}_3 - \boldsymbol{k}_4$, forms two triangles and the angle between them is the folding angle, $\Psi$. The $z$-axis is in the direction of $\boldsymbol{k}$ and the triangle $\boldsymbol{k} - \boldsymbol{k}_1 - \boldsymbol{k}_2 = 0$ is in the $x = 0$ plane. Here, $\pi - \Theta$ is the angle between $z$ - axis and $\boldsymbol{k}_2$ and $\Phi$ is the angle between $z$-axis and $\boldsymbol{k}_3$. These define the configuration of the quadrilateral. The viewing angles are $\theta$ and $\psi$. We then fix the shape by setting $k$, $s = k_2$, $t = k_3/k$, $\Theta$, $\Phi$, $\Psi$ implying $k_1 = k \sqrt{1 + s^2 + 2s \cos \theta}$, $k_4 = k \sqrt{1 + t^2 + 2t \cos \Phi}$.

\subsection{The odd and even parity trispectrum}
We have defined the parity-violating and parity-preserving parts of the trispectrum by
\begin{equation}
    T_{\mathrm{odd}} = \frac{1}{2}[T_g(\boldsymbol{k}_1,\boldsymbol{k}_2, \boldsymbol{k}_3,\boldsymbol{k}_4) - T_g(-\boldsymbol{k}_1,-\boldsymbol{k}_2, -\boldsymbol{k}_3,-\boldsymbol{k}_4)  ] \,,
\end{equation}
\begin{equation}
    T_{\mathrm{even}} = \frac{1}{2}[T_g(\boldsymbol{k}_1,\boldsymbol{k}_2, \boldsymbol{k}_3,\boldsymbol{k}_4) + T_g(-\boldsymbol{k}_1,-\boldsymbol{k}_2, -\boldsymbol{k}_3,-\boldsymbol{k}_4)  ].
\end{equation}
We expect $T_{\mathrm{odd}} = 0$, if we calculate the trispectrum with just the Newtonian kernels. However, more generally we expect it to be non-zero due to the terms in the relativistic kernels that only have odd powers of $\boldsymbol{n}$ and therefore odd powers of $\mu_i$. These odd powers arise due to the contributions from terms such as $\boldsymbol{v} \cdot \boldsymbol{{n}}$ and $\partial_r \delta$. We reiterate that these terms give rise to the parity-odd part of the trispectrum. Here, we explicitly calculate the even and odd parts and neglect terms that are $(\mathcal{H}/k)^2$ and higher. 

We now calculate the amplitude of the odd parity part relative to the even Newtonian part, with respect to the geometry above. 
We calculate the ratio $|T_{\mathrm{odd}}/T_{\mathrm{even}}|$ for two types of surveys. The amplitude for the two surveys is mainly different due to the magnification and the evolution biases. The two surveys that we have considered are Euclid-like near-infrared spectroscopic survey and SKA-like intensity mapping (IM) of 21cm neutral hydrogen radio emission. At $z = 1$, the Euclid-like survey has bias parameters $b_1 = 1.2, b_2 = -0.74, b_1' = - 1.6 \times 10^{-4} \mathrm{Mpc}^{-1}, b_e = -4, b_e' = 0, s = -0.95.$ (We set $\partial b_1 / \partial \mathrm{ln} L = 0 $ for simplicity.) Subsequently, the SKA-like survey has parameters $b_1 = 0.856, b_2 = -0.321, b_1' = -0.5 \times 10^{-4} \mathrm{Mpc}^{-1}, b_e = -0.5, b_e' =0, s = 1$. We ignore tidal bias and set the third-order bias to $0$ for simplicity. 
 
 \begin{figure} [h]
    \centering
    \includegraphics[width=0.89\linewidth]{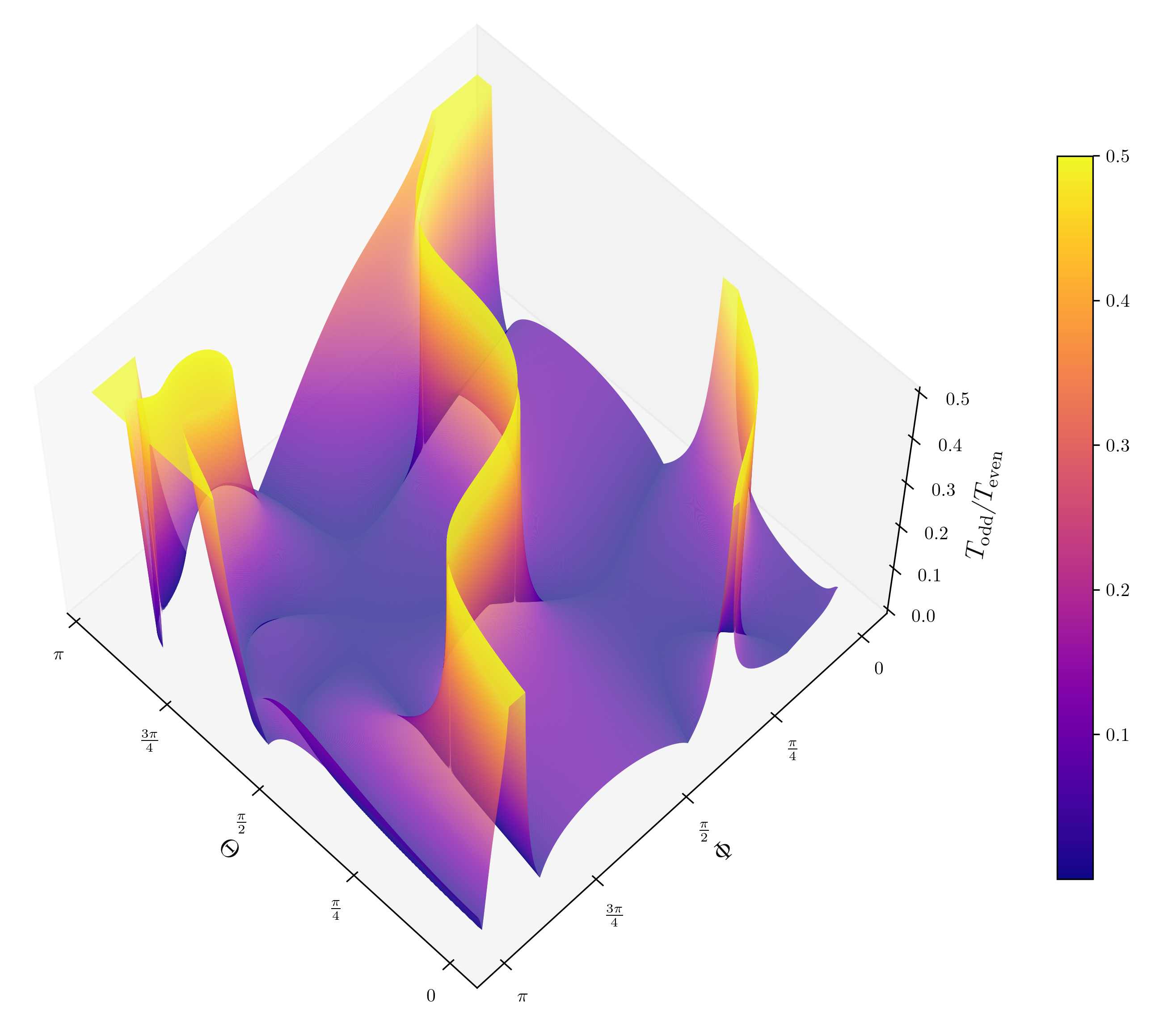}
    \caption{The ratio of the odd to even trispectrum $|T_{\rm{odd}}|/|T_{\rm{even}}|$ for an SKA-like survey as a function of configuration angles at $k = 0.01 h \mathrm{Mpc}^{-1}$. For all configurations, the relativistic corrections are $10\%$ or more. The bright yellow parts are where the even part changes sign. The geometry is $[ \Psi = 0.9\pi, k_2 = k, k_3 = 1.1k, \theta = 0, \phi = 0]$.}
    \label{fig:IM11}
\end{figure}

\begin{figure} [h]
    \centering
    \includegraphics[width=0.95\linewidth]{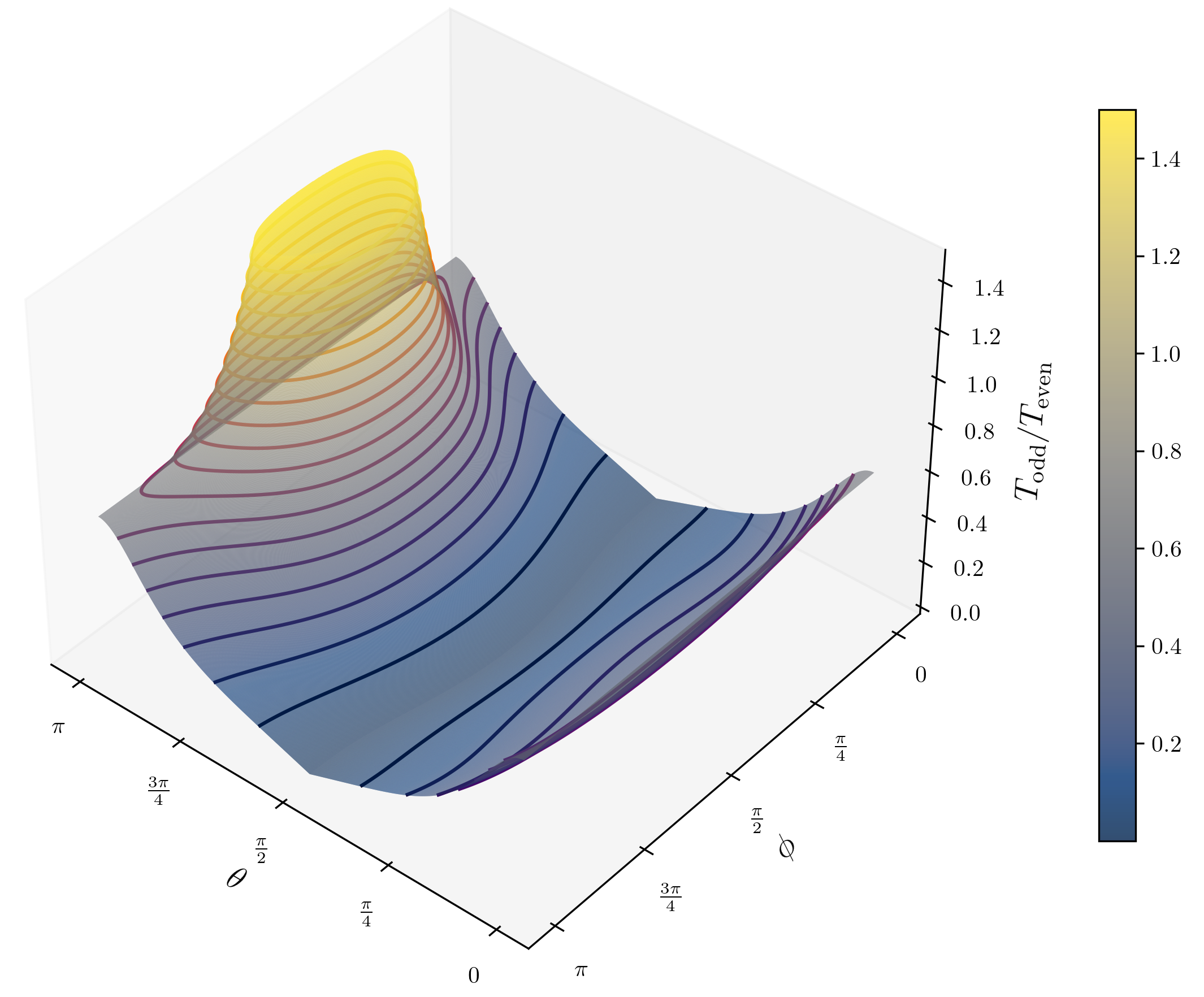}
    \caption{The ratio of the odd to even trispectrum, $|T_{\rm{odd}}|/|T_{\rm{even}}|$, for an SKA-like survey as a function of viewing angles at $k = 0.01 h \mathrm{Mpc}^{-1}$. For a fixed value of $\theta = \pi/2$, the relativistic corrections are $20 \%$. As $\theta$ varies, the corrections are more than $40\%$ and reaches as high as $80 \%$. The bright yellow part is where the even part changes sign. The geometry is $[\Theta = 0.9\pi, \Phi = 0.2\pi, \Psi = \pi, k_2 = k, k_3 = 1.1k]$.}
    \label{fig:IM31}
\end{figure}

Figs.~\ref{fig:IM11} and \ref{fig:IM31} display the ratio of the odd- to even-parity trispectrum for a SKA-like survey, with a fixed scale at $k = 0.01h \mathrm{Mpc}^{-1}$. In Fig.~\ref{fig:IM11}, the viewing angles and the folding angle is set to $\theta =\phi = 0$ and $\Psi = 0.9 \pi$ respectively. The colouring in Fig.~\ref{fig:IM11} is indicative of the amplitude of the corrections. For all combinations of configuration angles, we can see that the corrections are $10\%$ or more - i.e regions with colouring of purple. The peaks represented by yellow are regions where $T_{\mathrm{even}} =0$. 
In Fig.~\ref{fig:IM31}, the configuration angles are $\Theta = 0.9 \pi$, $\Phi = 0.2 \pi$ and the folding angle represents a flat quadrilateral, $\Psi = \pi$. For a fixed $\theta = \pi/2$, as $\phi$ varies, we have contributions of $20 \%$ from the parity-odd part. However, for all other values of $\theta$, as $\phi$ varies, the region gradually changes to grey which represents parity-odd contributions of $60\%$ to $80\%$. For values of $\pi/4 \lesssim \phi \lesssim  3\pi/4 $, at $\theta \approx \pi/2$, $T_{\mathrm{even}}$ changes sign. In both Figs.~\ref{fig:IM11} and \ref{fig:IM31}, we can see that the ratio significantly varies with both configuration and orientation angles. Comparing them, it is evident $T_{\mathrm{even}}$ changes sign more often in the combination of the configuration angles as compared to the viewing angles.

\subsection{Second and third order contributions}

In this section, we will discuss the second and third order contributions to the trispectrum separately. In (\ref{trispeccoeffs}), the terms that contribute at tree level are the combination of the first and third order terms and the first and second order terms. We define $T_{1113}$ and $T_{1122}$ as the third and second order contributions respectively. In this section, we have also considered the amplitudes for the same surveys as $z=1$. 

 \begin{figure} [h]
    \centering
    \includegraphics[width=0.99\linewidth]{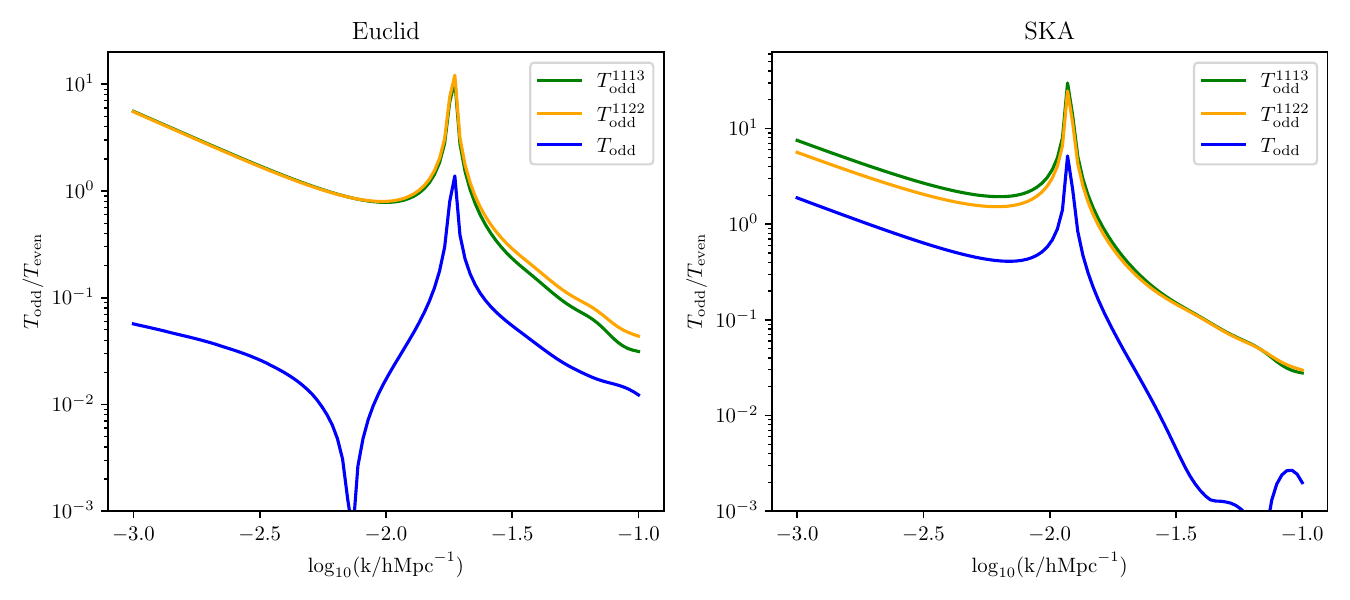}
    \caption{The ratio of $|T_{\mathrm{odd}}^{1113}|$ and $|T_{\mathrm{odd}}^{1122}|$ to the even trispectrum for a Euclid-like (left) and SKA-like (right) survey as a function of scale for a flattened configuration ($\Psi = \pi$). For both surveys, there are cancellations occurring between $T^{1113}$ and $T^{1122}$, with the amplitude of $T_{\rm{odd}}$ much smaller. For Euclid-like survey, $T^{1122}$ is the dominant contribution, with the trend opposite for the SKA-like survey. The geometry is $[\Theta = 0.9\pi, \Phi = 0.1\pi, \Psi = \pi, k_2 = 1.1k, k_3 = k, \theta = 0, \phi = 0]$.}
    \label{fig:IM3}

\end{figure}
  \begin{figure} [h]
    \centering
    \includegraphics[width=0.99\linewidth]{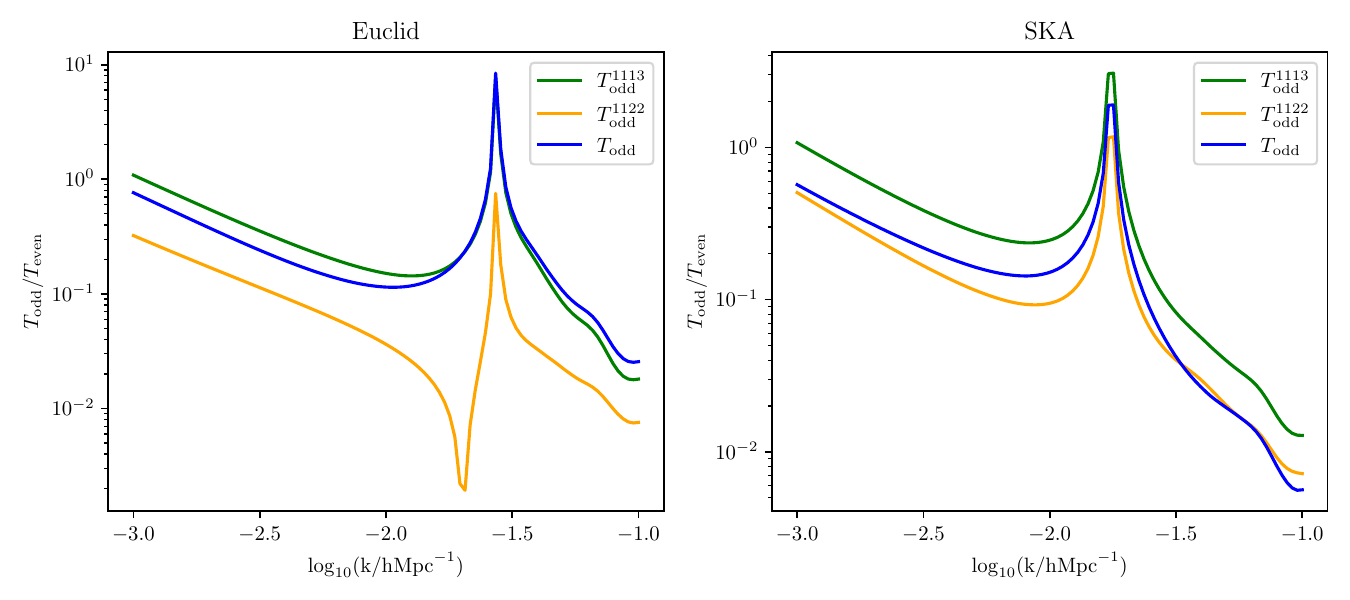}
    \caption{The ratio of $|T_{\mathrm{odd}}^{1113}|$ and $|T_{\mathrm{odd}}^{1122}|$ to the even trispectrum for a Euclid-like (left) and SKA-like (right) survey as a function of scale for a flattened configuration ($\Psi = \pi$). In this case, changing the viewing angles from Fig.~\ref{fig:IM3} reduces the cancellations that happen between $T^{1113}$ and $T^{1122}$. For both surveys, $T^{1113}$ is the dominant contribution. The geometry is $[\Theta = 0.9\pi, \Phi = 0.1\pi, \Psi = \pi, k_2 = 1.1k, k_3 = k, \theta = 0.7 \pi, \phi = 0.7 \pi]$.}
    \label{fig:IM4}
\end{figure}
\begin{figure} [h]
    \centering
    \includegraphics[width=0.99\linewidth]{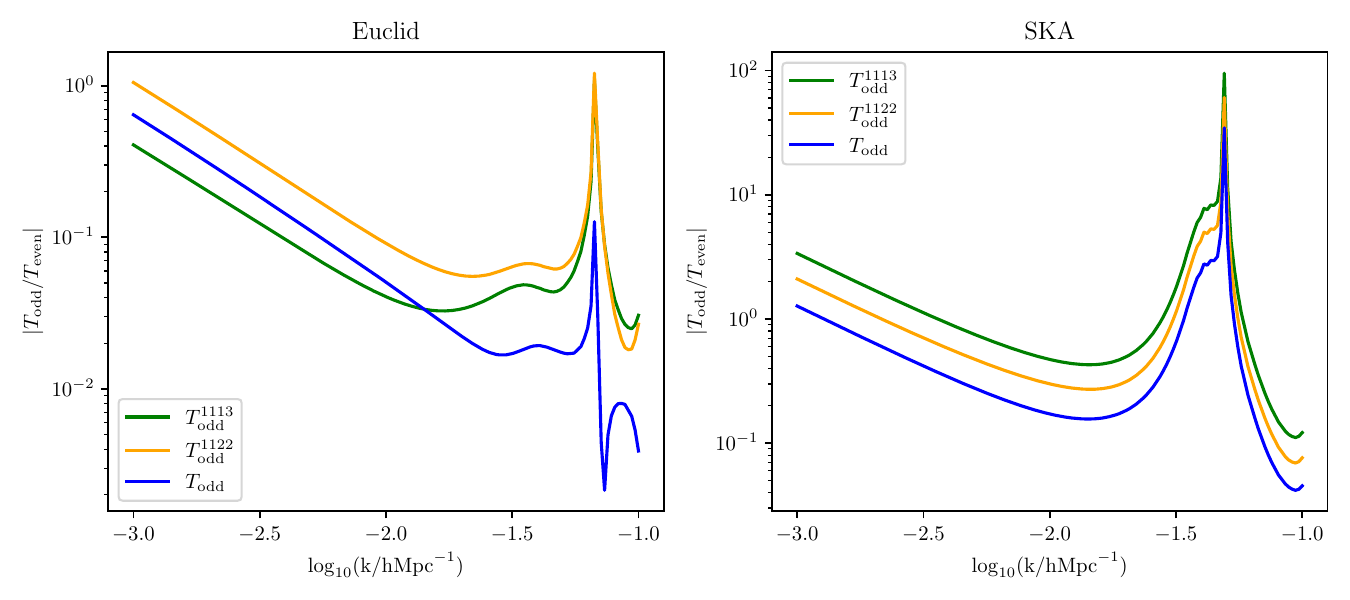}
    \caption{The ratio of $|T_{\mathrm{odd}}^{1113}|$ and $|T_{\mathrm{odd}}^{1122}|$ to the even trispectrum for a Euclid-like (left) and SKA-like (right) survey as a function of scale for a flattened configuration ($\Psi = \pi$). For the Euclid-like case, $T^{1122}$ contribution dominates for smaller values of $k$, while for the SKA-like case, $T^{1113}$ dominates. The geometry is $[\Theta = 0.1\pi, \Phi = 0.96\pi, \Psi = \pi, k_2 = 1.1k, k_3 = k, \theta = 0.2 \pi, \phi = 0.2 \pi]$.}
    \label{fig:IM5}
\end{figure}
\begin{figure} [h]
    \centering
    \includegraphics[width=0.99\linewidth]{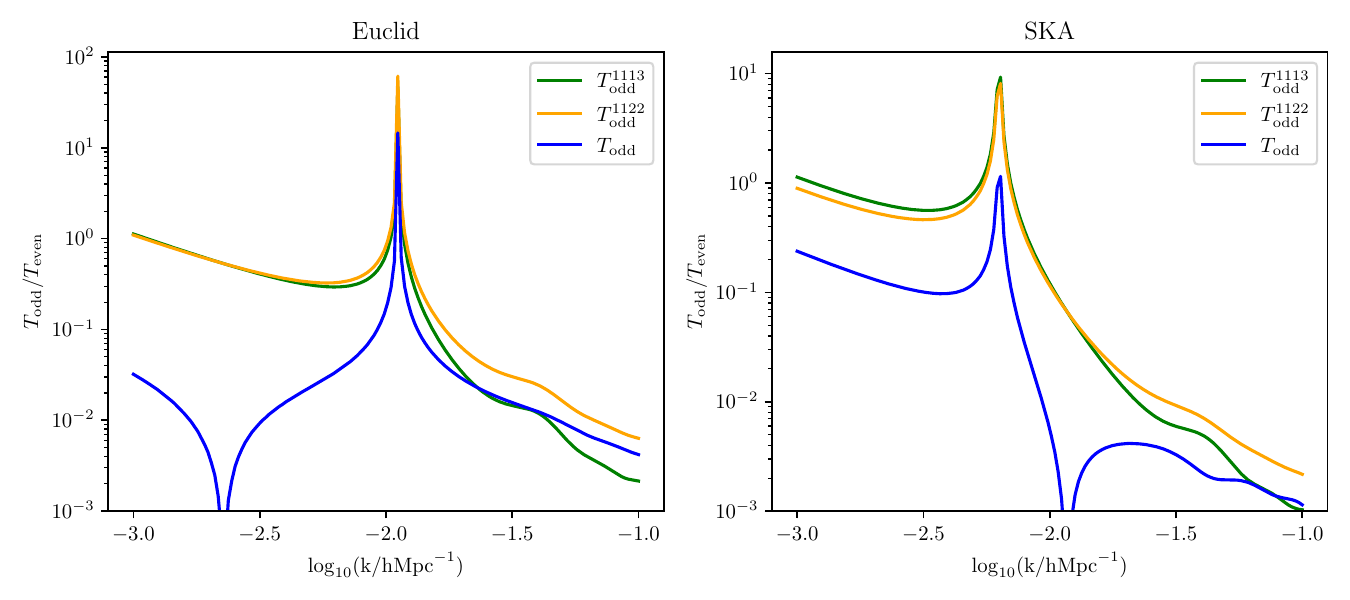}
    \caption{The ratio of $|T_{\mathrm{odd}}^{1113}|$ and $|T_{\mathrm{odd}}^{1122}|$ to the even trispectrum for a Euclid-like (left) and SKA-like (right) survey as a function of scale for a flattened configuration ($\Psi = \pi$). Compared to Fig.~\ref{fig:IM5}, cancellations in the contributions are more for the Euclid-like survey than the SKA-like survey. The geometry is $[\Theta = 0.2\pi, \Phi = 0.3\pi, \Psi = \pi, k_2 = 1.1k, k_3 = k, \theta = 0.2 \pi, \phi = 0.2 \pi]$.}
    \label{fig:IM6}
\end{figure}

Figs.~\ref{fig:IM3}, \ref{fig:IM4}, \ref{fig:IM5} and \ref{fig:IM6}  show the ratio of the third and second order odd contributions to the even part of the trispectrum as a function of scale. These are for a flattened configuration $(\Psi = \pi)$, with the tetrahedron lying on a plane. In Fig.~\ref{fig:IM3}, for the Euclid-like case the third and second contributions are similar in amplitude for values of $k < 0.01 h \mathrm{Mpc}^{-1}$. For values of $k > 0.03 h \mathrm{Mpc}^{-1}$, the second order contribution dominates over the third order. We see some cancellations happens between $T_{1113}$ and $T_{1122}$, as the total amplitude $T_\mathrm{odd}$ is small compared to the individual contributions. It is also interesting to note there are larger cancellations for $k < 0.01 h \mathrm{Mpc}^{-1}$. However, for the SKA-like case, the trend is the opposite. For values of $k < 0.01 h \mathrm{Mpc}^{-1}$, the amplitude of the third order contribution is more dominant. The cancellations that occur are also larger for values of $k > 0.03 h \mathrm{Mpc}^{-1}$. In Fig.~\ref{fig:IM3} and \ref{fig:IM4} we have kept the configuration and folding angles the same, only varying $\theta$ and $\psi$. We see that changing the viewing angles has a significant effect on which contribution dominates. For the Euclid-like case, the amplitude of the third order contributions is more dominant than the second order ones for all values of $k$. For these viewing angles, the same trend is followed by the SKA-like survey. We also see that for Fig.~\ref{fig:IM4}, the cancellation for both surveys is smaller compared to Fig.~\ref{fig:IM3}. So, we conclude that cancellations are dependent on the choice of the viewing angles.
  

In Figs.~\ref{fig:IM5} and \ref{fig:IM6}, we plotted the odd and even ratios varying the configuration angles. For the Euclid-like cases, we see that the cancellations in Fig.~\ref{fig:IM6} are much more compared to Fig.~\ref{fig:IM5}, while for the SKA-like case, the difference is less noticeable. In Fig.~\ref{fig:IM5}, the second order contributions dominate over the third order for large values of $k$, although for the same configuration, the SKA-like survey has the opposite trend. In Fig.~\ref{fig:IM6}, for values of $k < 0.01 h \mathrm{Mpc}^{-1}$, there are significant cancellations as the total $T_\mathrm{odd}$ contribution has a smaller amplitude than the third or the second order contribution, for the Euclid-like survey. For the SKA-like survey, both Figs.~\ref{fig:IM5} and \ref{fig:IM6} don't have significant cancellations. As a result, we conclude that even for the same configuration angles, varying the surveys have a significant impact on the amplitudes of the individual and the total contributions. When varying configurations, cancellations between different terms also alter significantly. 

 \begin{figure} [h]
    \centering
    \includegraphics[width=0.99\linewidth]{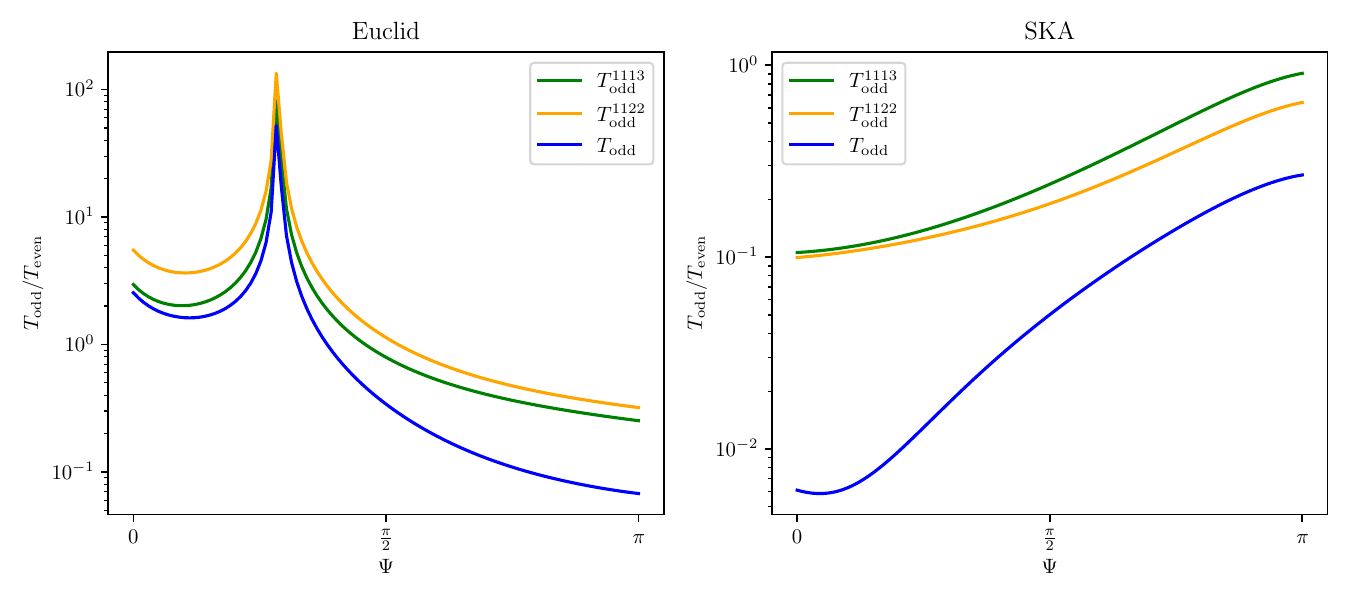}
    \caption{The ratio of $|T_{\mathrm{odd}}^{1113}|$ and $|T_{\mathrm{odd}}^{1122}|$ to the even trispectrum for a Euclid-like (left) and SKA-like (right) survey as a function of the folding angle $\Psi$, at $k = 0.01h \mathrm{Mpc^{-1}}$. For the Euclid-like case, $T_{\mathrm{even}} =0$ at $\Psi \sim 0.98$. For the SKA-like case, the cancellations in the contributions increases signifcantly as $\Psi \lesssim \pi/2$. The geometry is $[\Theta = 0.8\pi, \Phi = 0.1\pi, k_2 = 1.1k, k_3 = k, \theta = 0.7 \pi, \phi = 0.7 \pi]$.}
    \label{fig:IM7}
\end{figure}

  \begin{figure} [h]
    \centering
    \includegraphics[width=0.99\linewidth]{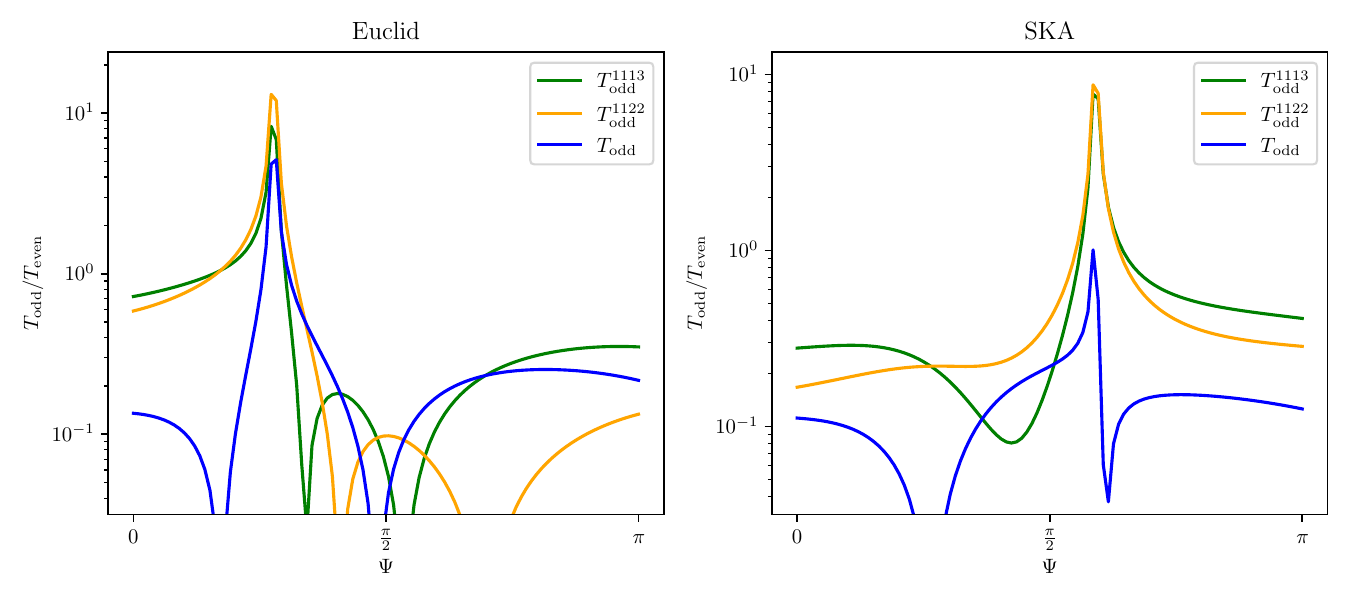}
    \caption{The ratio of $|T_{\mathrm{odd}}^{1113}|$ and $|T_{\mathrm{odd}}^{1122}|$ to the even trispectrum for a Euclid-like (left) and SKA-like (right) survey as a function of the folding angle $\Psi$, at $k = 0.01h \mathrm{Mpc^{-1}}$. As we changed the configuration angles from Fig.~\ref{fig:IM7}, we also have $T_{\mathrm{odd}}=0$ for several values of $\Psi$. For the SKA-like case, $T^{1113}$ and $T^{1122}$ are exactly equal to each other for $\Psi \approx \pi/4$. The geometry is $[\Theta = 0.2\pi, \Phi = 0.9\pi, k_2 = 1.1k, k_3 = k, \theta = 0.7 \pi, \phi = 0.7 \pi]$.}
    \label{fig:IM8}
\end{figure}

In Figs.~\ref{fig:IM7} and \ref{fig:IM8}, we have now plotted the ratio of the third and second order odd contributions to the even part of the trispectrum as a function of $\Psi$, for different configuration angles. We see that for the same viewing and configuration angles, plotting different surveys can alter the amplitude of the third and second order contributions significantly. For the Euclid-like case, the third order contribution dominates for all values $\Psi$, while the second order contribution dominates for the SKA-like case. In the Euclid-like case, for $\Psi \sim 0.98$, we have $T_{\mathrm{even}} = 0$.  This is not the case for the SKA-like survey and the total amplitude of $T_{\mathrm{odd}}$ is significantly less than the individual contributions for values of $\Psi \lesssim \pi/2$.  From Fig.~\ref{fig:IM7} to Fig.~\ref{fig:IM8}, we have changed the configuration angles which alters the shape of the graphs. For the SKA-like case, as $\Psi \approx \pi/4$, the $T_{\mathrm{odd}}^{1113}$ and $T_{\mathrm{odd}}^{1122}$ changes signs and are exactly equal to each other, while this property is not seen in Fig.~\ref{fig:IM7}.

\subsection{First and second order kernels}
 
\begin{figure}[t] 
\begin{centering}
\includegraphics[width=0.49\textwidth]{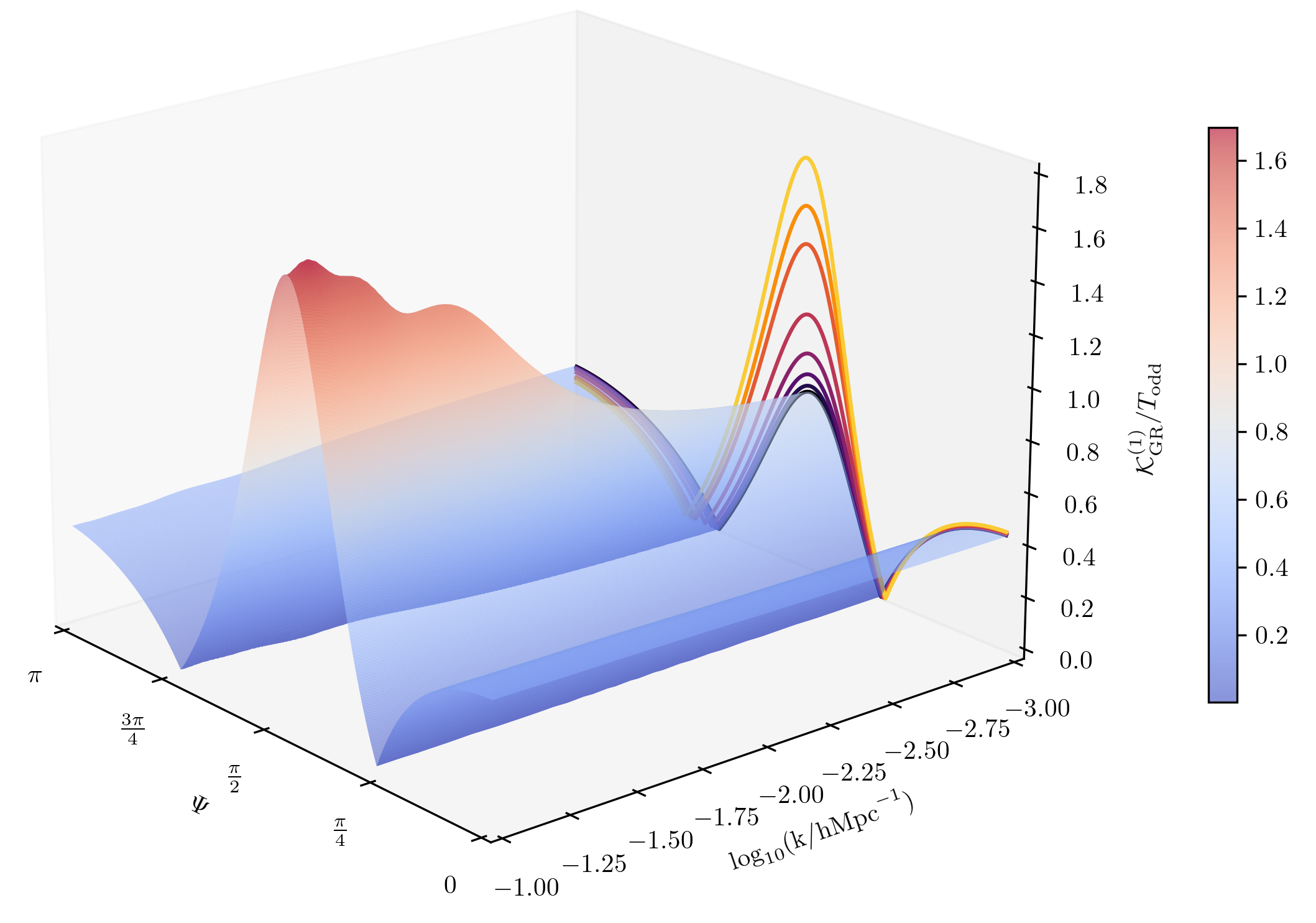}
\includegraphics[width=0.49\textwidth]{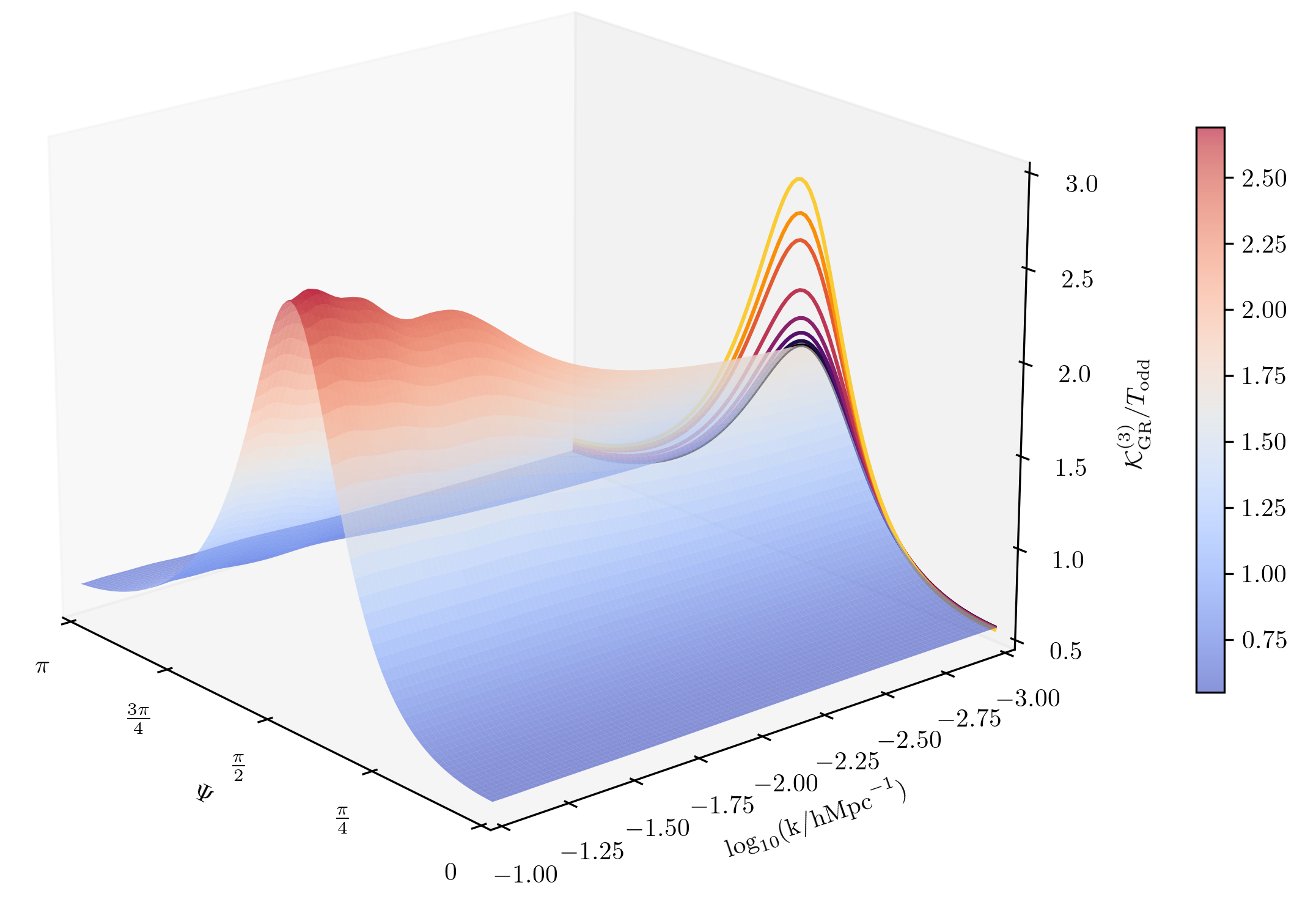}
\caption{The ratio of the first and third order kernels to the odd trispectrum for a SKA-like survey as a function of scale and folding angle $\Psi$. For $\mathcal{K}_{\mathrm{GR}}^{(1)}$, the ratio goes to zero for values of $\Psi = \pi/4$ and $3\pi/4$, which is not seen for $\mathcal{K}_{\mathrm{GR}}^{(3)}$. The geometry is $[\Theta = 0.2\pi, \Phi = 0.96\pi, k_2 = 1.1k, k_3 = k, \theta = 0.7 \pi, \phi = 0.7 \pi]$.}
\label{IM11133}
\end{centering}
\end{figure}

\begin{figure}[t]
\begin{centering}
\includegraphics[width=0.49\textwidth]{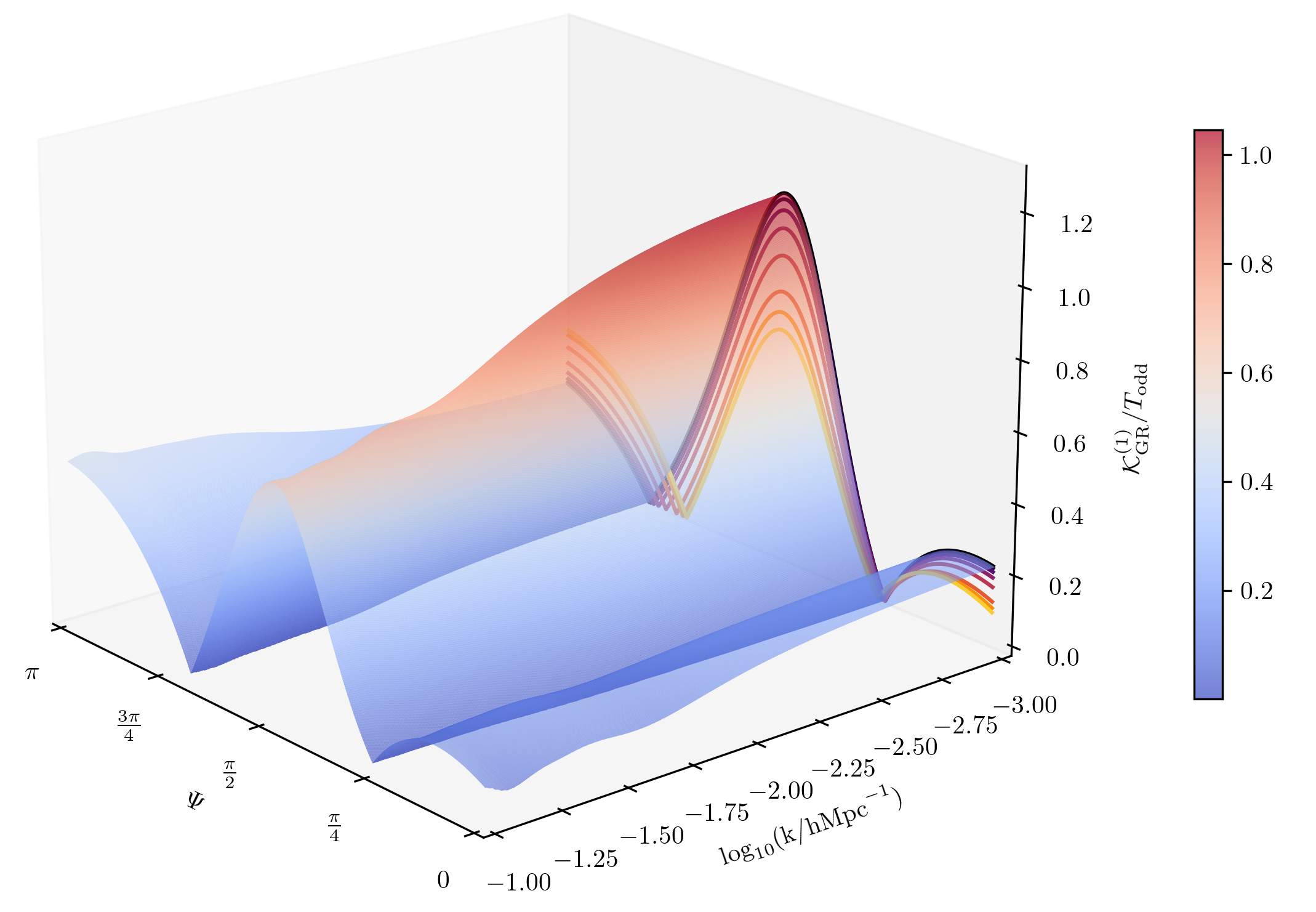}
\includegraphics[width=0.49\textwidth]{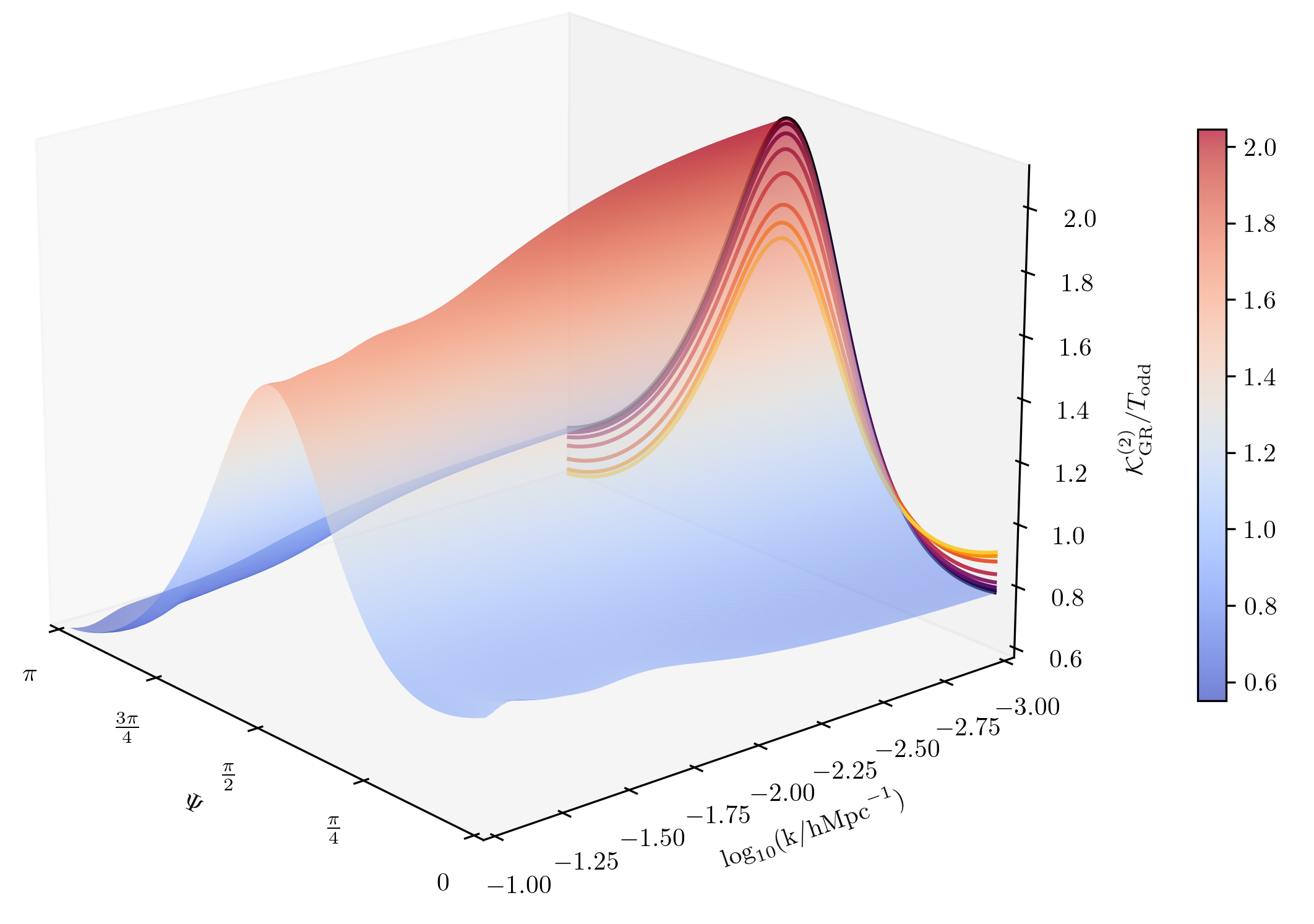}
\caption{The ratio of the first and second order kernels to the odd trispectrum for a SKA-like survey as a function of scale and folding angle $\Psi$. For the second order contribution, the ratio for both the kernels increases as smaller values of $k$, i.e larger scales. The geometry is $[\Theta = 0.2\pi, \Phi = 0.96\pi, k_2 = 1.1k, k_3 = k, \theta = 0.7 \pi, \phi = 0.7 \pi]$.}
\label{IM11222}
\end{centering}
\end{figure}

 The general relativistic contributions to the $T_{1113}$ terms are the $\mathcal{K}^{(1)}_{\mathrm{GR}}$ and $\mathcal{K}^{(3)}_{\mathrm{GR}}$ order kernels. Similarly, the general relativistic contributions to the $T_{1122}$ terms are the $\mathcal{K}^{(1)}_{\mathrm{GR}}$ and $\mathcal{K}^{(2)}_{\mathrm{GR}}$ order kernels.
 We now calculate the amplitude of the individual kernels relative to the full third and second order contribution. 
\begin{figure} [h]
    \centering
    \includegraphics[width=0.99\linewidth]{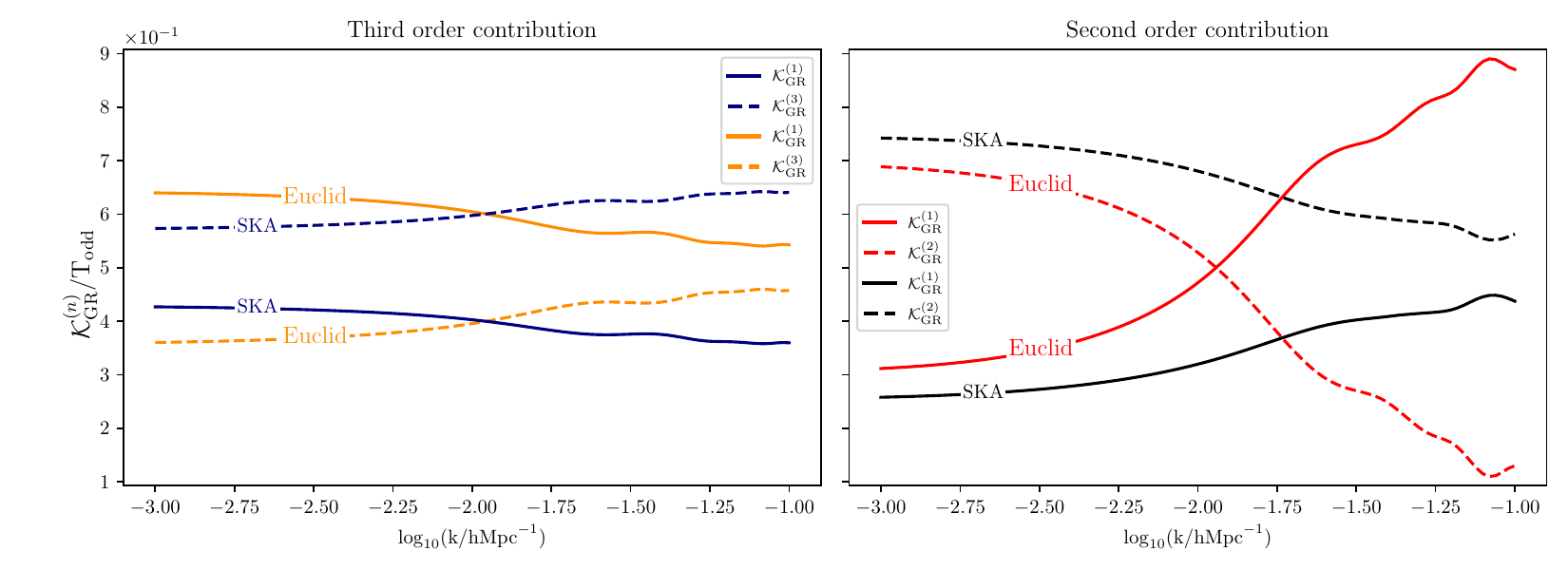}
    \caption{The ratio of the first and third order kernels to the odd trispectrum for a Euclid and SKA-like survey as a function of scale for a flattened configuration ($\Psi = \pi$). For the third order and the second contribution, for the SKA-like case, $\mathcal{K}_{\mathrm{GR}}^{(1)}$ dominates over $\mathcal{K}_{\mathrm{GR}}^{(3)}$. For the Euclid-like case, in the third order contribution, $\mathcal{K}_{\mathrm{GR}}^{(1)}$ dominates.  The geometry is $[\Theta = 0.2\pi, \Phi = 0.96\pi, k_2 = 1.1k, k_3 = k, \theta = 0.7 \pi, \phi = 0.7 \pi]$. }
    \label{fig:IM1}
\end{figure}
 Figs.~\ref{fig:IM1}, \ref{IM11133}, and \ref{IM11222} display the ratio of the first, second and third order kernels to the odd-parity part of the trispectrum for a flattened configuration. In Fig.~\ref{fig:IM1}, on the left-hand side of the plot, we have ratios corresponding to the $T_{1113}$ contribution to the odd trispectrum. For an Euclid-like survey, for all values of $k$, $\mathcal{K}^{(1)}_{\mathrm{GR}}$ contribution to $T_{\mathrm{odd}}$ dominates over $\mathcal{K}^{(3)}_{\mathrm{GR}}$. For values of $k < 0.014 h \mathrm{Mpc^{-1}}$, the first order contribution is significantly more dominant than the third order contribution, compared to the other values of $k$. For the SKA-like survey, for all values of $k$, $\mathcal{K}^{(3)}_{\mathrm{GR}}$ contribution to $T_{\mathrm{odd}}$ dominates over $\mathcal{K}^{(1)}_{\mathrm{GR}}$. Here, for values of $k > 0.014 h \mathrm{Mpc^{-1}}$, the third order contribution significantly dominates the first order contribution, when compared to the other values of $k$. For the right-hand side of the graph, we have ratios corresponding to the $T_{1122}$ contribution. For an Euclid-like survey, for values of $k > 0.014 h \mathrm{Mpc^{-1}}$, $\mathcal{K}^{(2)}_{\mathrm{GR}}$ kernel dominates over the $\mathcal{K}^{(1)}_{\mathrm{GR}}$ kernel. For all other values of $k$, the first order kernel dominates over the third. For the SKA-like survey, $\mathcal{K}^{(2)}_{\mathrm{GR}}$ dominates for all values of $k$. 
 

Figs.~\ref{IM11133} and \ref{IM11222} investigate the relationship between the folding angle, scale and the ratio $\mathcal{K}^{(n)}_{\mathrm{GR}}/T_{\mathrm{odd}}$, for $n =1,2,3$. In Fig.~\ref{IM11133}, both $\mathcal{K}^{(1)}_{\mathrm{GR}}$ and $\mathcal{K}^{(3)}_{\mathrm{GR}}$, follow the same trend with varying $k$. The Baryonic Acoustic Oscillations are clearly visible below the equality scales for $\Psi = \pi/2$. For $\mathcal{K}^{(1)}_{\mathrm{GR}}$, for values of $\Psi = \pi/4$ and $3\pi/4$, $\mathcal{K}^{(1)}_{\mathrm{GR}} = 0$, while for $\mathcal{K}^{(3)}_{\mathrm{GR}}$, there are no values of $\Psi$ where the ratio goes to $0$. In Fig.~\ref{IM11222}, for both $\mathcal{K}^{(1)}_{\mathrm{GR}}$ and $\mathcal{K}^{(2)}_{\mathrm{GR}}$, the ratio increases as we go to larger scales - i.e smaller values of $k$. Following the same trend as Fig.~\ref{IM11133}, $T_{\mathrm{even}} = 0$ for the same values of $\Psi$ for the $\mathcal{K}^{(1)}_{\mathrm{GR}}$ kernel. 

\subsection{The odd and even parity trispectrum against redshift}

In this section, we investigate how the odd parity trispectrum varies with changing redshift. We plot $|T_{\mathrm{odd}}/T_{\mathrm{even}}|$ as a function of redshift as a fixed $k = 0.01h\mathrm{Mpc}^{-1}$, for different viewing angles. 
\begin{figure}[t]
\begin{centering}
\includegraphics[width=0.49\textwidth]{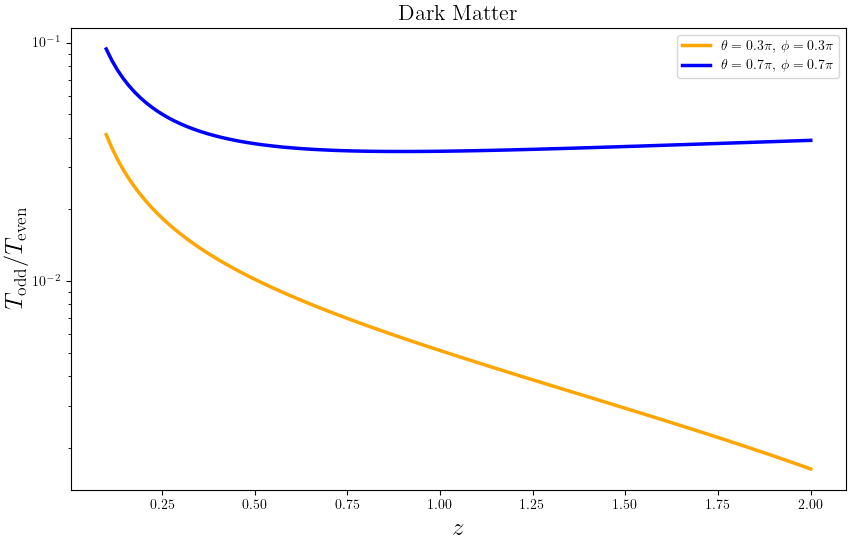}
\includegraphics[width=0.49\textwidth]{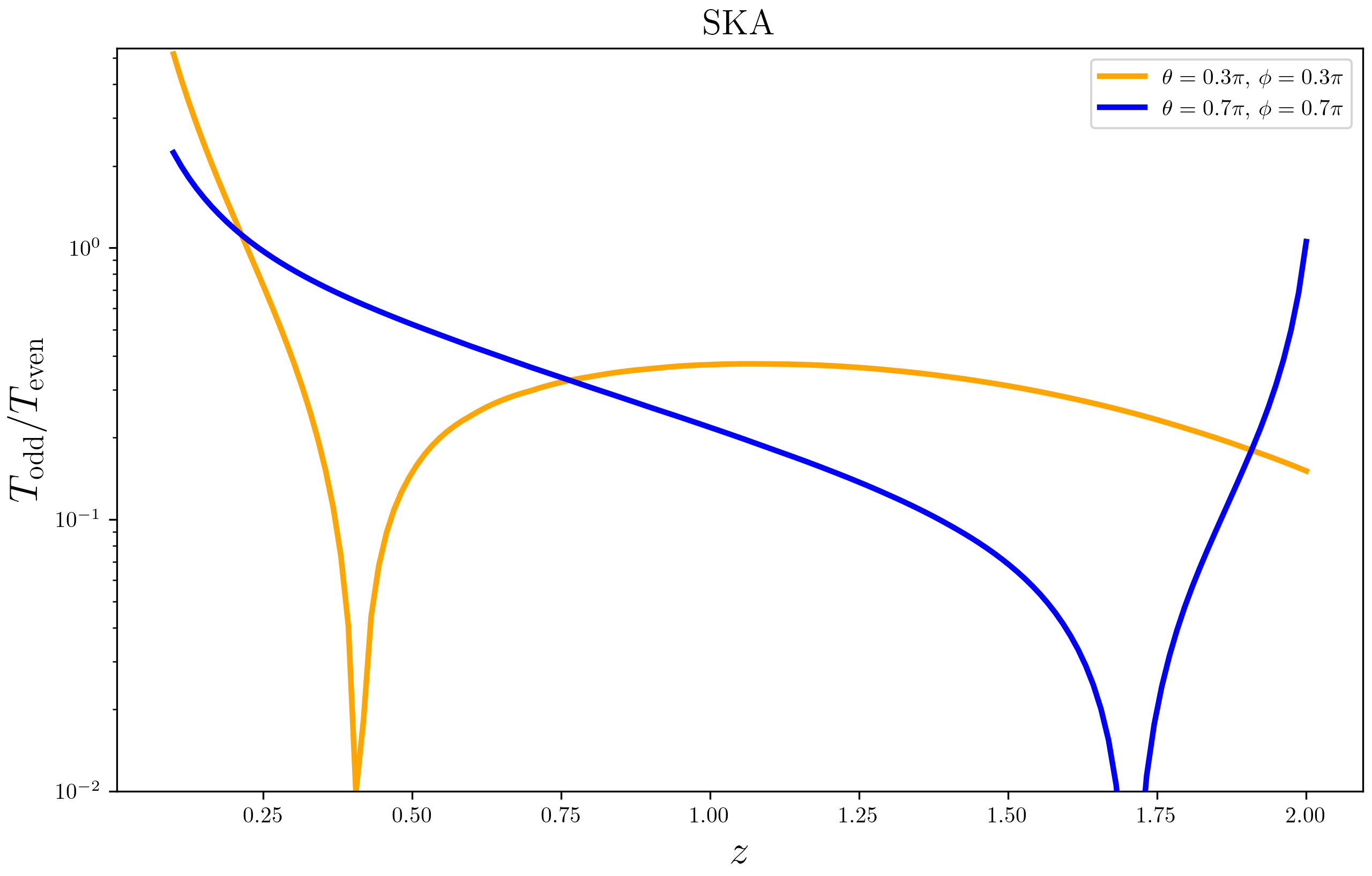}
\caption{The ratio of the odd to even trispectrum, $|T_{\mathrm{odd}}/T_{\mathrm{even}}|$, for dark matter (left) and SKA-like survey (right) as a function of redshift, $z$, and two different viewing angles. The geometry is $[\Theta = 0.2\pi, \Phi = 0.8\pi, k_2 = 1.1k, k_3 = k, \Psi = \pi]$.}
\label{redshift}
\end{centering}
\end{figure}

In order to estimate the expected redshift dependence of $|T_{\mathrm{odd}}/T_{\mathrm{even}}|$, we note that most of the time dependence cancels in the ratio. Each contribution to this will be of the form $\sim \mathcal{K}^{(n)}_{\mathrm{GR}}/\mathcal{K}^{(n)}_{\mathrm{N}}$ where $n=1,2,3$. Since there is no time dependence in $\mathcal{K}^{(n)}_{\mathrm{N}}$ except through the bias and $f$ factors, the main time dependence is in $\mathcal{K}^{(n)}_{\mathrm{GR}}\sim\mathcal{H}$ together with evolution and magnification bias as the main contributions. Therefore, we can expect the effects of the odd-parity part to be larger or smaller at higher redshifts depending on the tracer of interest.


In Fig.~\ref{redshift}, we plot the dark matter distribution (left) with galaxy bias, $b=1$, and the evolution and magnification biases set to zero. On the right, we have plotted the same configuration for an SKA-like survey. The galaxy bias is given by 
\bea
&b_1 = 0.750 + 0.087z + 0.019z^2. \\ \notag
\eea
The evolution and magnification biases are taken from Table 2 in \cite{Maartens_2021}.
For the dark matter distribution, the ratios are straight lines varying with redshift, as they only depend on $f$ and $\mathcal{H}$. For the SKA-like survey, the terms coupled with the biases also come into play and displays a different trend. We also see that the evolution with redshift is sensitive to the choice of viewing angles. For the SKA-like survey, for $\theta = \phi = 0.3\pi$, the ratio has a slight increase but gets suppressed as we go to higher redshift. For $\theta = \phi = 0.7\pi$, we have the ratio decreasing for values of $0.3 \lesssim z \lesssim 1.5$. This highlights the importance of the dependence on evolution and magnification biases. 

\subsection{Summary of Numerical Results}

We have explored the different parameters that we used to evaluate the trispectrum. These include the scale $k$, configuration angles, $\Theta, \Phi$, viewing angles, $\theta, \phi$, and folding angle, $\Psi$. For both the configuration angles, we conclude that the parity-odd corrections are $10 \%$ or more at equality scales. Changing the viewing angles, we conclude for some values of $\theta$ and $\phi$, the corrections can go up to $80 \%$. 
We also investigated the ratio of the individual third, $T^{1113}$, and second order contributions, $T^{1122}$, to the even-parity part. There are lots of interesting cancellations going on in the individual contributions, which results in a smaller amplitude of the total ratio, $T_{\mathrm{odd}}/T_{\mathrm{even}}$. For the same configuration and viewing angles, changing the surveys switches the dominating contribution. As an overall trend, we see the odd contributions increase for larger scales, i.e smaller values of $k$. We have also looked into contributions of the different kernels $\mathcal{K}_{\mathrm{GR}}^{(n)}$, where $n =1,2,3$. When we consider $T_{\mathrm{odd}}$, we not only have the contributions which are purely $\mathcal{K}^{(n)}_{\mathrm{GR}}$, but also permutations where $\mathcal{K}^{(n)}_{\mathrm{GR}}$ is coupled with $\mathcal{K}^{(n)}_{\mathrm{N}}$ contributions. For the second order contribution, we see that $\mathcal{K}^{(1)}_{\mathrm{GR}}/T_{\mathrm{even}}$ and $\mathcal{K}^{(2)}_{\mathrm{GR}}/T_{\mathrm{even}}$ increases with decreasing $k$. 

\section{Conclusion}

Several studies have been carried out to see whether the universe is parity-odd. If the late-time universe is indeed parity-odd, this could provide a window into new physics which has occurred during inflation. It can also help constrain many inflation theories, including multi-field inflation \cite{coulton2023signaturesparityviolatinguniverse,Zhu:2024wme,hou2024baryonacousticoscillationsilluminate}. Whilst this has been looked at in data concerning the CMB and gravitational waves, there has been less analysis searching for parity violation in the scalar sector. 

In this paper, we look into the effect that relativistic redshift space distortion can have on the trispectrum. These relativistic effects are responsible for producing a parity-odd trispectrum, even though the primordial trispectrum may be parity preserving. These relativistic effects have been previously calculated in the multi tracer power spectrum and the bispectrum, where they are responsible for an imaginary part of the N-point spectra. For the trispectrum, these effects are significantly more complicated and require a third order analysis. The leading contribution is suppressed by $\mathcal{H}/k$ in comparison to the standard RSD terms and is invariant under the transformation $\boldsymbol{k} \rightarrow -\boldsymbol{k}$. We have explored the parameter space of the trispectrum, which contains the viewing angles and the configuration angles. We have seen for most configurations, these corrections cannot be neglected and for equality scales, they are around $10 \%$ or more. We have also seen that there are interesting cancellations that occur between the third and second order contributions which result in an overall lower signal. The corrections also significantly vary for different surveys. We have explicitly calculated the corrections for Euclid- and SKA-like surveys. However, for both the surveys, the overall trend is the same and implies that these relativistic corrections cannot be neglected. Although the standard RSD terms are not parity violating and the monopole of the trispectrum does not have this parity violation, the dipole has a significant contribution from these relativistic terms. With surveys probing higher fractions of the sky and larger redshifts, these contributions become dominant and should be taken into account in any further analysis of the trispectrum. One has to be very careful to remove any possible contaminants when measuring the observed trispectrum. Even though it is unlikely that \cite{Hou_2023} has detected these effects, as they have averaged over all lines of sight, it is important to take these effects into account in future to ensure the analysis is accurate. 

With new upcoming surveys, we will have a huge volume of large scale structure data. Further analysis can be done to compute a signal-to-noise ratio for the trispectrum. This will tell us how well we can detect and constrain the relativistic effects in the trispectrum. Joint analysis can also be done with the bispectrum and the trispectrum to get tighter constraints for the relativistic effects corrections. Alongside looking at the parity-odd part of the trispectrum, we can also use the statistic for a more theoretically accurate assessment of primordial non-Gaussianity.

\appendix

\section{Fourier transform kernel and coefficient} \label{Fourtrans}
 In order to calculate the relativistic kernel, we use the same convention as \ref{Fourconv}. In third order, there are many coupling terms that we take into account. We define the radial and transverse component of the velocity as
 
 \begin{align} \label{dersradtrans}
    &v_{||}(\boldsymbol{k}) = i \mu k v(\boldsymbol{k})\,, \\
    &v_\perp^a(\boldsymbol{k}) = i \big[{k}^a - \mu k\,{n}^a\big] v(\boldsymbol{k})\quad \mbox{or}\quad 
\bm{v}_\perp=i\,\bm{k}_\perp\,v(\bm k).
\end{align}
We also use Fourier transformation of a product, which results in a convolution in Fourier space: 
\bea 
   h(\boldsymbol{k}) = \int \frac{\mathrm{d^3} k_1}{(2 \pi)^3} \frac{\mathrm{d^3} k_2}{(2 \pi)^3} f(\boldsymbol{k}_1) g(\boldsymbol{k}_2) (2 \pi)^3 \delta^D(\boldsymbol{k}_1 + \boldsymbol{k}_2 - \boldsymbol{k})
\eea
We have given examples of typical terms calculated in Fourier space below.  The terms that contribute to $v^{(3)}(\boldsymbol{x}) \delta_g^{(3)}(\boldsymbol{x})$ at third order are $\frac{1}{2} v^{(1)}(\boldsymbol{x}) \delta_g^{(2)}(\boldsymbol{x})$ and $ \frac{1}{2}v^{(2)}(\boldsymbol{x}) \delta_g^{(1)}(\boldsymbol{x})$. We have a factor of $1/2$ due to our convention in (\ref{conv}). In Fourier space, $v^{(1)}(\boldsymbol{x}) \delta_g^{(2)}(\boldsymbol{x})$ can be expressed as
\begin{align}
    v^{(1)}_{||}(\boldsymbol{x}) \delta_g^{(2)}(\boldsymbol{x}) = \int \frac{\d^3 k}{(2 \pi)^3} e^{i\boldsymbol{k}\cdot\boldsymbol{x}} \big[v^{(1)}_{||}\delta^{(2)}_g \big] (\boldsymbol{k})\,,
\end{align}
\begin{align}
    \Big[v^{(1)}_{||} \delta_g^{(2)} \Big](\boldsymbol{k}) &= \int {\d^3 x} e^{-i\boldsymbol{k}\cdot\boldsymbol{x 
    }} v^{(1)}_{||}(\boldsymbol{x}) \delta_g^{(2)}(\boldsymbol{x})\,, \\ \notag
    &= \int \d^3 x \int \frac{\d^3k_3}{(2 \pi)^3} \int \frac{\d^3k_4}{(2 \pi)^3} \Big[ v^{(1)}_{||}(\boldsymbol{x}) \delta_g^{(2)}(\boldsymbol{x})\Big] e^{-i \boldsymbol{k} \cdot \boldsymbol{x}} e^{i \boldsymbol{k}_3 \cdot \boldsymbol{x}} e^{i \boldsymbol{k}_4 \cdot \boldsymbol{x}}\, \\ \notag
    &= \int \frac{\d^3k_3}{(2 \pi)^3} \int \frac{\d^3k_4}{(2 \pi)^3} \Big[ v^{(1)}_{||}(\boldsymbol{x}) \delta_g^{(2)}(\boldsymbol{x})\Big] (2 \pi)^3 \delta^D(\boldsymbol{k}_3 + \boldsymbol{k}_4 - \boldsymbol{k}).
\end{align}
We can then express the variables
\begin{align}
    &v_{||}^{(1)}(\boldsymbol{k}_3) = if \frac{\mu_3}{k_3} \mathcal{H} \delta^{(1)}(\boldsymbol{k}_3) \,,   \\
    & \delta_g^{(2)} (\boldsymbol{k}_4) = b_1 \delta^{(2)} (\boldsymbol{k}_4) + b_2 \delta^2(\boldsymbol{k}_4).
\end{align}
By substituting everything in, this leads to 
\begin{align}
    \Big[ v^{(1)} \delta_g^{(2)}\Big](\boldsymbol{k}) = \int \frac{\d^3 k_1}{(2 \pi)^3} \frac{\d^3 k_2}{(2 \pi)^3} \frac{\d^3 k_3}{(2 \pi)^3} \mathcal{F} \Big[v^{(1)} \delta_g^{(2)}\Big] \delta^{(1)}(\boldsymbol{k}_1) \delta^{(1)}(\boldsymbol{k}_2) \delta^{(1)}(\boldsymbol{k}_3) (2\pi)^3 \delta^D(\boldsymbol{k}_1 + \boldsymbol{k}_2 + \boldsymbol{k}_3 - \boldsymbol{k})
\end{align}
where the kernel is - 
\begin{align}
    \mathcal{F} \Big[v^{(1)} \delta_g^{(2)}\Big] = if\mathcal{H} \frac{\mu_3}{k_3} \Big[ b_2 F_2(\boldsymbol{k}_1, \boldsymbol{k}_2) + b_2 \Big]. 
\end{align}
Another example, where there are only first order terms, is $\partial_r(v^2_{||} \delta)$. Following the same steps as above
\begin{align}
    \partial_r\Big[v_{||}^2(\boldsymbol{x})\delta (\boldsymbol{x}) \Big] = \int \frac{\d^3 k}{(2 \pi)^3} e^{i\boldsymbol{k} \cdot \boldsymbol{x}}\Big[ \partial_r(v_{||}^2 \delta) \Big] (\boldsymbol{k}).
\end{align}
\begin{align}
    \Big[\partial_r(v_{||}^2 \delta)\Big](\boldsymbol{k}) &= \int \d^3 x e^{-i\boldsymbol{k} \cdot \boldsymbol{x}} \partial_r[v_{||} (\boldsymbol{x}) \delta(\boldsymbol{x})], \\ \notag
    & = \int \d^3 x \int \frac{\d^3 k_1}{(2 \pi)^3} \int \frac{\d^3 k_2}{(2 \pi)^3} \int  \frac{\d^3 k_3}{(2 \pi)^3} \partial_r\Big[v_{||}^2(\boldsymbol{x})\delta(\boldsymbol{x}) \Big] e^{-i\boldsymbol{k} \cdot \boldsymbol{x}} e^{i\boldsymbol{k}_1 \cdot \boldsymbol{x}} e^{i\boldsymbol{k}_2 \cdot \boldsymbol{x}} e^{i\boldsymbol{k}_3 \cdot \boldsymbol{x}} \\ \notag
    & = \int \frac{\d^3 k_1}{(2 \pi)^3} \int \frac{\d^3 k_2}{(2 \pi)^3} \int  \frac{\d^3 k_3}{(2 \pi)^3} \partial_r\Big[v_{||}^2(\boldsymbol{x})\delta(\boldsymbol{x}) \Big] (2 \pi)^3 \delta^D(\boldsymbol{k}_1 + \boldsymbol{k}_2 + \boldsymbol{k}_3 - \boldsymbol{k})
\end{align}
We can then use equation \ref{dersradtrans} to get the following
\begin{align}
    \Big[ \partial_r(v^2_{||}\delta) \Big](\boldsymbol{k}) = \int \frac{\d^3 k_1}{(2 \pi)^3} \frac{\d^3 k_2}{(2 \pi)^3} \frac{\d^3 k_3}{(2 \pi)^3} \mathcal{F} \Big[\partial_r(v^2_{||}\delta)\Big] \delta^{(1)}(\boldsymbol{k}_1) \delta^{(1)}(\boldsymbol{k}_2) \delta^{(1)}(\boldsymbol{k}_3) (2\pi)^3 \delta^D(\boldsymbol{k}_1 + \boldsymbol{k}_2 + \boldsymbol{k}_3 - \boldsymbol{k})\,,
\end{align}
where the kernel is
\begin{align}
    \mathcal{F}\Big[ \partial_r(v_{||}^2 \delta)(\boldsymbol{k}) \Big] = -i \frac{\mu_1 \mu_2}{k_1 k_2} f^2 \mathcal{H}^2 (\mu_1 k_1 + \mu_2 k_2 + \mu_3 k_3).
\end{align}

\section{The coefficients in the Newtonian kernel $\mathcal{K}^{(3)}_{\mathrm{N}}$} \label{newtcoeff}
 \begin{center}
 \scalebox{1}{
 \begin{tabular}{
 c c c} 
 \hline
 Term & Fourier kernel $\mathcal{F}$ & Coefficient \\ [0.5ex] 
 \hline\hline 
 \\
 $\delta^{(3)}$ & $ F_3(\boldsymbol{k}_1, \boldsymbol{k}_2, \boldsymbol{k}_3)$ & $b_1$ 
 \\ \\
 $\delta \delta^{(2)}$ & $ F_2(\boldsymbol{k}_1, \boldsymbol{k}_2)$ & $b_2$ 
 \\ \\
  $\partial_rv_{||}^{(3)}$ & $\mu_4^2 f \mathcal{H} G_3(\boldsymbol{k}_1, \boldsymbol{k}_2, \boldsymbol{k}_3)$ & $\mathcal{H}^{-1}$  
  \\ \\
$\partial_r\big(v_{||}^{(1)} \delta_g^{(2)}\big)$ & $\mu_3 (\mu_1 k_1 + \mu_2 k_2 + \mu_3 k_3) f \mathcal{H} \big( b_1 F_2(\boldsymbol{k}_1, \boldsymbol{k}_2) +  b_2 \big)/k_3$& $\mathcal{H}^{-1}$  
\\ \\
$\partial_r\big(v_{||}^{(2)} \delta_g^{(1)}\big)$ & $(\mu_1 k_1 + \mu_2 k_2) (\mu_1 k_1 + \mu_2 k_2 + \mu_3 k_3) b_1 f \mathcal{H} G_2(\boldsymbol{k}_1, \boldsymbol{k}_2)/|\boldsymbol{k}_1 + \boldsymbol{k}_2|^2$ & $\mathcal{H}^{-1}$  
\\ \\
$\partial_r\big(v_{||}^{(2)} \partial_rv_{||}^{(1)}\big)$ & $\mu_3^2(\mu_1 k_1 + \mu_2 k_2) (\mu_1 k_1 + \mu_2 k_2 + \mu_3 k_3) f^2 \mathcal{H}^2 G_2(\boldsymbol{k}_1, \boldsymbol{k}_2) / |\boldsymbol{k}_1 + \boldsymbol{k}_2|^2 $& $\mathcal{H}^{-2}$ 
\\ \\
$\partial_r\big(v_{||}^{(1)} \partial_rv_{||}^{(2)}\big)$ &$ \mu_3 (\mu_1k_1 + \mu_2k_2)^2(\mu_1k_1 + \mu_2k_2+ \mu_3 k_3) f^2 \mathcal{H}^2 G_2(\boldsymbol{k}_1, \boldsymbol{k}_2)/k_3|\boldsymbol{k}_1 + \boldsymbol{k}_2|^2$ & $\mathcal{H}^{-2}$  
\\ \\ 
 $\partial^3_r v_{||}^{(3)}$ &$ (\mu_1k_1 + \mu_2k_2 + \mu_3 k_3)^3 \mu_1  \mu_2  \mu_3  f^3 \mathcal{H}^3/k_1 k_2 k_3 $ & $\frac{1}{6} \mathcal{H}^{-3}$  
 \\ \\
$\partial^2_r(\delta_g v_{||}^2)$ & $\mu_2 \mu_3 (\mu_1 k_1 + \mu_2k_2 + \mu_3 k_3)^2 b_1 f^2 \mathcal{H}^2/k_2 k_3$ & $\frac{1}{2} \mathcal{H}^{-2}$ \\ [1ex] 
 \hline
\end{tabular}}
\end{center}

\newpage

\section{The coefficients in the GR kernel $\mathcal{K}^{(3)}
_{\mathrm{GR}}$} \label{relcoeff}

 \hspace{-1.2cm}
 \scalebox{0.7}{
 \begin{tabular}{
 c c c} 
  \hline \\
  Term & Fourier kernel $\mathcal{F}$ & Coefficient \\ [0.5ex] 
  \hline\hline
  \\ 
 $v_{||}^{(3)}$ & $-i \mu_4 f \mathcal{H} G_3(\boldsymbol{k}_1, \boldsymbol{k}_2, \boldsymbol{k}_3) /k_4$ & ${\mathcal{H}'}/\mathcal{H}^2 + (2 - 5s)/\mathcal{H}r + 5s - b_e$  
 \\ \\ 
 $v_{||}^{(1)} \delta_g^{(2)}$ & $- i  \mu_3 f\mathcal{H} (b_1 F_2(\boldsymbol{k}_1 , \boldsymbol{k}_2) + b_2) / {k_3}$ & ${\mathcal{H}'}/\mathcal{H}^2 + (2 - 5s)/\mathcal{H}r + 5s - b_e$
  \\ \\ 
 $v_{||}^{(2)} \delta_g^{(1)} $ & $ - i (\mu_1k_1 + \mu_2 k_2) f \mathcal{H} b_1G_2(\boldsymbol{k}_1, \boldsymbol{k}_2)/|\boldsymbol{k}_1 + \boldsymbol{k}_2|^2$& ${\mathcal{H}'}/\mathcal{H}^2 + (2 - 5s)/\mathcal{H}r + 5s - b_e$  
 \\ \\ 
  $v_{||}^{(1)} \partial_r v_{||}^{(2)}$ & $- i  \mu_3 (\mu_1 k_2 + \mu_2 k_2)^2 f^2 \mathcal{H}^2 G_2(\boldsymbol{k}_1, \boldsymbol{k}_2)/k_3|\boldsymbol{k}_1 + \boldsymbol{k}_2|^2 $  & $(1 + 3 {\mathcal{H}'}/\mathcal{H}^2 + (4-5s)/\mathcal{H}r + 5s - 2 b_e)\mathcal{H}^{-1}$   \\ \\ 
$v_{||}^{(2)} \partial_r v_{||}^{(1)}$ & $- i  \mu^2_3(\mu_1k_1 + \mu_2 k_2) f^2 \mathcal{H}^2 G_2(\boldsymbol{k}_1, \boldsymbol{k}_2)/|\boldsymbol{k}_1+ \boldsymbol{k}_2|^2 $  &  $(1 + 3 {\mathcal{H}'}/\mathcal{H}^2 + (4-5s)/\mathcal{H}r + 5s - 2 b_e)\mathcal{H}^{-1}$   \\ \\ 
$v_{||}^{(1)} {\delta}_g'^{(2)}$ & $ - i\mu_3 f^2\mathcal{H}^2 (b_1 F_2(\boldsymbol{k}_1 , \boldsymbol{k}_2) + b_2) / {k_3}$& $- \mathcal{H}^{-1}$  \\ \\ 
$v_{||}^{(2)} {\delta}_g'^{(1)}$ & $-i (\mu_1k_1 + \mu_2 k_2) f^2 \mathcal{H}^2 b_1G_2(\boldsymbol{k}_1, \boldsymbol{k}_2)/|\boldsymbol{k}_1 + \boldsymbol{k}_2|^2 $& $- \mathcal{H}^{-1}$ 
\\ \\ 
$v_{\perp}^{(1)} \partial_{\perp} v_{||}^{(2)}$ & $i (\mu_1k_1 + \mu_2k_2)  [\boldsymbol{k}_3 \cdot (\boldsymbol{k}_1 + \boldsymbol{k}_2) - \mu_3 k_3 (\mu_1k_1 + \mu_2 k_2)] f^2 \mathcal{H}^2 G_2(\boldsymbol{k}_1, \boldsymbol{k}_2)/k_3^2 |\boldsymbol{k}_1 + \boldsymbol{k}_2|^2$  & $2\mathcal{H}^{-1}$  
\\ \\ 
$v_{\perp}^{(2)} \partial_{\perp} v_{||}^{(1)}$ & $i f^2 \mathcal{H}^2 \mu_3[(\boldsymbol{k}_1 + \boldsymbol{k}_2) \cdot \boldsymbol{k}_3 - (\mu_1k_1 + \mu_2 k_2)\mu_3k_3]G_2(\boldsymbol{k}_1, \boldsymbol{k}_2)/k_3|\boldsymbol{k}_1 + \boldsymbol{k}_2|^2$ & $2\mathcal{H}^{-1}$  \\ \\ 
$\psi^{(1)} \partial^2_r v_{||}^{(2)}$ &  $-3i \Omega_m (\mu_1k_1 + \mu_2 k_2)^3 f \mathcal{H}^3 G_2(\boldsymbol{k_}1, \boldsymbol{k}_2)/2k_3^2|\boldsymbol{k}_1 + \boldsymbol{k}_2|^2$ & $\mathcal{H}^{-2}$  \\ \\ 
$\psi^{(2)} \partial^2_r v_{||}^{(1)}$ &  $-3i \Omega_m \mu_3^3 k_3 f \mathcal{H}^3 F_2(\boldsymbol{k}_1, \boldsymbol{k}_2)/2|\boldsymbol{k}_1 + \boldsymbol{k}_2|^2$ & $\mathcal{H}^{-2}$  
\\ \\ 
$\psi^{(1)} \partial_r \delta_{g}^{(2)}$ & $-3i  \Omega_m (\mu_1k_1 + \mu_2 k_2) \mathcal{H}^2(b_1 F_2(\boldsymbol{k}_1, \boldsymbol{k}_2) +  b_2)/2k_3^2 $  & $\mathcal{H}^{-1}$ 
\\ \\
$\psi^{(2)} \partial_r \delta_{g}^{(1)}$ & $-3i \Omega_m \mu_3 k_3 \mathcal{H}^2 b_1 F_2(\boldsymbol{k}_1, \boldsymbol{k}_2)/2|\boldsymbol{k}_1 + \boldsymbol{k}_2|^2$  & $\mathcal{H}^{-1}$ 
\\ \\
 $v_{||}^{(2)} \partial^2_r \psi^{(1)}$ & $-3i \Omega_m  \mu_3^2(\mu_1 k_1 + \mu_2 k_2) f \mathcal{H}^3 G_2(\boldsymbol{k}_1, \boldsymbol{k}_2) /2|\boldsymbol{k}_1 + \boldsymbol{k}_2|^2$ & $-\mathcal{H}^{-2}$ \\ \\
 $ v_{||}^{(1)} \partial^2_r \psi^{(2)}$ & $-3i \Omega_m \mu_3 (\mu_1k_1 + \mu_2k_2)^2 f \mathcal{H}^3 F_2(\boldsymbol{k}_1, \boldsymbol{k}_2) / 2k_3|\boldsymbol{k}_1 + \boldsymbol{k}_2|^2$  & $-\mathcal{H}^{-2}$  
 \\ \\
 $ \partial_r v_{||} \partial^2_r v_{||} \psi$ & $-3i \Omega_m \mu_1^2 \mu_2^3 k_2 f^2 \mathcal{H}^4/2k_3^2$ &  $3\mathcal{H}^{-3}$ 
 \\ \\
 $ \partial_r v_{||} v_{||} \partial^2_r \psi$ &  $-3i \Omega_m \mu_1^2 \mu_2 \mu_3^2 f^2 \mathcal{H}^4/2k_2$ & $-3\mathcal{H}^{-3}$ 
 \\ \\
 $ v_{||} \partial^3_r v_{||} \psi$ & $3i \Omega_m f^2 \mathcal{H}^4 \mu_1 \mu_2^4 k_2^2/2k_1 k_3^2$  & $\mathcal{H}^{-3}$ 
 \\ \\ 
 $ v_{||} v_{||} \partial^3_r \psi$ & $3i \Omega_m \mu_1 \mu_2 \mu_3^3 k_3 f^2 \mathcal{H}^4 /2k_1 k_2$  & $- \mathcal{H}^{-3}$ 
 \\ \\
 $ \psi \partial_r^2(v_{||} \delta_g)$ & $-3i \Omega_m \mu_2(\mu_2 k_2 + \mu_3 k_3)^2 f \mathcal{H}^3/2k_1^2 k_2$  & $ \mathcal{H}^{-2}$ 
 \\ \\
 $ \partial_r^2 \psi (v_{||} \delta_g)$ &  $3i \Omega_m f \mathcal{H}^3 \mu_1^2 \mu_2 b_1 /2k_2$ & $- \mathcal{H}^{-2}$ 
 \\ \\
 $ \partial_r^2(v_{||}^3)$ & $-i \mu_1 \mu_2 \mu_3 (\mu_1 k_1 + \mu_2 k_2 + \mu_3 k_3)^2 f^3 \mathcal{H}^3/ k_1 k_2 k_3$  & $(1 + 3{\mathcal{H}'}/\mathcal{H}^2 + 3/(\mathcal{H}r) - 3 b_e/2) / 3\mathcal{H}^2$ 
 \\ \\
 $ \partial_r(v_{||}^2 \delta)$ & $-i\mu_1 \mu_2 (\mu_1k_1 + \mu_2 k_2 + \mu_3 k_3) f^2 \mathcal{H}^2 /k_1 k_2$  & $(1 + 3{\mathcal{H}'}/\mathcal{H}^2 + 4/(\mathcal{H}r) - 2 b_e) / 2\mathcal{H}$ 
 \\ \\
 $ \partial_r(v_{||}^2 {\delta}')$ & $ - i \mu_1 \mu_2 (\mu_1 k_1 + \mu_2 k_2 + \mu_3 k_3) f^3 \mathcal{H}^3/k_1 k_2$  & $-1/\mathcal{H}^2$ 
 \\ \\
 $ \partial_r^2(v_{||} v^{\perp} v_{\perp} )$ & $-i \mu_1 f^3 \mathcal{H}^3 (\mu_1 k_1 + \mu_2 k_2 + \mu_3 k_3)^2 (\boldsymbol{k}_2 \cdot \boldsymbol{k}_3 - \mu_2k_2 \mu_3 k_3)/k_1 k_2^2 k_3^2$  & $-1/2\mathcal{H}^2$ 
 \\ \\
 $ \partial_r(\delta v^{\perp} v_{\perp} )$ & $-i f^2 \mathcal{H}^2 (\mu_1 k_1 + \mu_2 k_2 + \mu_3 k_3) (\boldsymbol{k}_2 \cdot \boldsymbol{k}_3 - \mu_2 k_2 \mu_3 k_3)/k_2^2 k_3^2 $  & $-1/\mathcal{H}$ 
 \\ \\
 $ v^2 \partial_r \delta $ &  $-i \mu_3 k_3 (\boldsymbol{k}_2 \cdot \boldsymbol{k}_3) f^2 \mathcal{H}^2/ k_1^2 k_2^2$ & $1/2\mathcal{H}$ 
 \\ \\ 
 $ \partial_{||}(\partial_{\perp} v_{||} v^{\perp} v_{||}) $ &  $if^3 \mathcal{H}^3 \mu_1 \mu_3 (\mu_1 k_1 + \mu_2 k_2 + \mu_3 k_3)(\boldsymbol{k}_1 \cdot \boldsymbol{k}_2 - \mu_1 k_1 \mu_2 k_2)/k_1k_2^2k_3$ & $2/\mathcal{H}^2$ \\[1ex]
\hline \\
\end{tabular}}
 \clearpage

\acknowledgments

RM is supported by the South African Radio Astronomy Observatory and the National Research Foundation (grant no. 75415).

\bibliographystyle{JHEP}
\bibliography{ref}



\end{document}